\documentclass[reqno,11pt]{amsart}
\usepackage[utf8]{inputenc}
\usepackage{graphicx}
\usepackage{amscd}
\usepackage{slashed}
\usepackage{amssymb}
\usepackage{esint}
\usepackage[mathscr]{eucal}
\numberwithin{equation}{section}
\allowdisplaybreaks[1]

\newcommand{\SetFigFont}[3]{}

\title[The Fermionic Projector in Space-Times of Infinite Lifetime]{A Non-Perturbative Construction of the
Fermionic Projector on Globally Hyperbolic Manifolds II -- Space-Times of Infinite Lifetime}

\author[F.\ Finster]{Felix Finster}
\address{Fakult\"at f\"ur Mathematik \\ Universit\"at Regensburg \\ D-93040 Regensburg \\ Germany}
\email{finster@ur.de}
\author[M.\ Reintjes]{Moritz Reintjes \\ \\ December 2013}
\address{IMPA - Instituto Nacional de Matem{\'a}tica Pura e Aplicada \\ Rio de Janeiro, Brasil}
\email{moritzreintjes@gmail.com}
\thanks{M.R.\ is supported by the Deutsche Forschungsgemeinschaft (DFG), RE 3471/2-1.}

\newtheorem{Def}{Definition}[section]
\newtheorem{Thm}[Def]{Theorem}
\newtheorem{Prp}[Def]{Proposition}
\newtheorem{Lemma}[Def]{Lemma}

\newtheorem{Corollary}[Def]{Corollary}

\newcommand{\Thanks}{\vspace*{.5em} \noindent \thanks}
\newcommand{\beq}{\begin{equation}}
\newcommand{\eeq}{\end{equation}}
\newcommand{\Proof}{\begin{proof}}
\newcommand{\QED}{\end{proof} \noindent}

\newcommand{\la}{\langle}
\newcommand{\ra}{\rangle}
\newcommand{\bra}{\mathopen{<}}
\newcommand{\ket}{\mathclose{>}}
\newcommand{\Sl}{\mathopen{\prec}}
\newcommand{\Sr}{\mathclose{\succ}}

\newcommand{\C}{\mathbb{C}}
\newcommand{\R}{\mathbb{R}}
\newcommand{\1}{\mbox{\rm 1 \hspace{-1.05 em} 1}}

\newcommand{\N}{\mathbb{N}}

\newcommand{\nuslsh}{\slashed{\nu}}
\renewcommand{\H}{\mathscr{H}}

\newcommand{\uslsh}{\slashed{u}}

\newcommand{\bep}{\begin{pmatrix}}
\newcommand{\enp}{\end{pmatrix}}

\newcommand{\Dir}{{\mathcal{D}}}
\newcommand{\D}{{\mathscr{D}}}

\newcommand{\B}{{\mathscr{B}}}

\newcommand{\Lin}{\text{\rm{L}}}
\newcommand{\Cisc}{C^\infty_{\text{\rm{sc}}}}
\newcommand{\Cisco}{C^\infty_{\text{\rm{sc}},0}}

\DeclareMathOperator{\diag}{diag}

\DeclareMathOperator{\supp}{supp}

\newcommand{\p}{\mathfrak{p}}
\newcommand{\Sig}{\mathscr{S}}
\newcommand{\scrM}{\mycal M}
\newcommand{\scrN}{\mycal N}

\DeclareFontFamily{OT1}{rsfso}{}
\DeclareFontShape{OT1}{rsfso}{m}{n}{ <-7> rsfso5 <7-10> rsfso7 <10-> rsfso10}{}
\DeclareMathAlphabet{\mycal}{OT1}{rsfso}{m}{n}

\begin{document}

\begin{abstract}
The previous functional analytic construction of the fermionic projector on globally
hyperbolic Lorentzian manifolds is extended to space-times of infinite lifetime.
The construction is based on an analysis of
families of solutions of the Dirac equation with a varying mass parameter.
It makes use of the so-called mass oscillation property which implies that
integrating over the mass parameter generates decay of the Dirac wave functions at infinity.
We obtain a canonical decomposition of the solution space of the massive Dirac equation
into two subspaces, independent of observers or the choice of
coordinates. The constructions are illustrated in the examples of ultrastatic space-times and
de Sitter space-time.
\end{abstract}

\maketitle

\tableofcontents

\section{Introduction}
In the recent paper~\cite{finite}, the fermionic projector was constructed non-perturbatively
in a space-time of finite lifetime. In the present paper, we extend the construction to
space-times of infinite lifetime. In order to introduce the problem, we begin with the simplest possible
example: the Minkowski vacuum. We thus consider
the vacuum Dirac equation
\[ (i \gamma^j \partial_j - m)\, \psi(x) = 0 \]
in Minkowski space~$(\scrM, \la .,. \ra)$. On solutions of the Dirac equation one has the scalar product
\beq \label{sprodMin}
( \psi | \phi)_m := \int_{\R^3} (\overline{\psi} \gamma^0 \phi)(t, \vec{x})\: d^3x
\eeq
(which by current conservation is independent of~$t$; here~$\overline{\psi} \equiv \psi^\dagger \gamma^0$
is the so-called adjoint spinor).
Moreover, on wave functions with suitable decay at infinity (which do not need to be
solutions of the Dirac equation), we can introduce a Lorentz invariant inner product
by integrating over space-time,
\beq \label{stipMin}
\bra \psi|\phi \ket = \int_\scrM \overline{\psi(x)} \phi(x) \: d^4x \:.
\eeq
In~\cite{finite} we proceeded by representing the space-time inner product~\eqref{stipMin}
with respect to the
scalar product~\eqref{sprodMin} as
\beq \label{rel2}
\bra \psi | \phi \ket = ( \psi | \Sig \phi )_m
\eeq
with a signature operator~$\Sig$. Then the positive and negative spectral
subspaces of the operator~$\Sig$ gave the desired splitting of the solution space into
two subspaces.
Unfortunately, in Minkowski space an identity of the form~\eqref{rel2} makes no mathematical sense.
Namely, the right side of~\eqref{rel2} is defined only if~$\psi$ and~$\phi$ are solutions of the Dirac equation.
But on solutions, the left side of~\eqref{rel2} is ill-defined
because the time integral in~\eqref{stipMin} will in general diverge.

Our method to overcome this problem is to work with families of solutions
with a varying mass parameter. This can be understood most easily if one takes the
spatial Fourier transform,
\[ \psi(t, \vec{x}) = \int_{\R^3} \frac{d^3k}{(2 \pi)^3} \:\hat{\psi}(t, \vec{k})\: e^{i \vec{k} \vec{x}} \:. \]
Then a family of solutions has the representation
\[ \hat{\psi}_m(t, \vec{k}) = c_+(\vec{k},m) \: e^{-i t \omega(\vec{k},m)} + c_-(\vec{k},m)
\: e^{i t \omega(\vec{k},m)} \]
with suitable spinor-valued coefficients~$c_\pm$, where we set~$\omega(\vec{k},m):=\sqrt{|\vec{k}|^2+m^2}$.
For a suitable class of solutions (for example families which are smooth and compactly supported
in~$m$ and~$\vec{x}$), the coefficients~$c_\pm$ are smooth functions of~$m$.
If~$m \neq 0$, the derivative~$\partial_m \omega(\vec{k}, m)$ is non-zero, implying that the
phase factors~$e^{\pm i t \omega(m,\vec{k})}$ oscillate in~$m$.
The larger~$t$ is chosen, the faster these phase factors
oscillate if~$m$ is varied. This implies that if we integrate over~$m$ by setting
\beq \label{ppsiMin}
(\p \hat{\psi})(t, \vec{k}) = \int_I \hat{\psi}_m(t, \vec{k})\: dm \:,
\eeq
we obtain destructive interference of a
superposition of waves with different phases (here~$I \subset \R \setminus \{0\}$ is an interval
containing the support of~$c_\pm(\vec{k},.)$).
If~$t$ is increased, the integrand oscillates faster in~$m$, so that the integral becomes smaller.
We thus obtain decay in time.
This intuitive picture that oscillations in the mass parameter give rise to decay for large times
is made mathematically precise by the {\em{mass oscillation property}}.
We shall prove that, using the mass oscillation property, one can give~\eqref{rel2} a mathematical
meaning by inserting suitable mass integrals,
\beq \label{aim}
\bra \p \psi | \p \phi \ket = \int_I ( \psi_m | \Sig_m \phi_m )_m \: dm \:.
\eeq
We thus obtain a family of bounded linear operators~$\Sig_m$. For any fixed mass~$m$,
the positive and negative spectral subspaces of the operator~$\Sig_m$ give rise to a canonical
decomposition of the solution space into two subspaces.

It it the main purpose of this paper to make such ideas and methods applicable
in the general setting of globally hyperbolic manifolds.
After the preliminaries in Section~\ref{secprelim}, we 
begin by stating the most general assumptions on the Dirac operator in space-time
under which mass oscillations can be studied, referred to as the {\em{weak mass oscillation
property}} (Section~\ref{secwMOP}). In this setting, the operators~$\Sig_m$ cannot be defined for fixed~$m$,
but only the combination~$\Sig_m \,dm$ is defined as an operator-valued measure.
In Section~\ref{secsMOP} we introduce stronger assumptions (the {\em{strong mass oscillation property}})
which ensure that the operators~$\Sig_m$ are bounded operators which are uniquely defined for
any~$m \in I$. We point out that we state the mass oscillation properties purely in terms
of the solution spaces of the Dirac equation. This has the advantage that we do not need to
make any assumptions on the asymptotic behavior of the metric at infinity.
The strong mass oscillation property also makes it possible to define the
fermionic projector as an integral operator with a distributional kernel.

In the last two sections we illustrate the abstract constructions by simple examples.
Section~\ref{secultra} is devoted to the Dirac operator in ultrastatic space-times, possibly involving
an arbitrary static magnetic field. We find that in this ultrastatic situation, the positive and negative
spectral subspaces of the operator~$\Sig_m$ coincide precisely with the solutions of positive
and negative frequency. We thus obtain agreement with the ``frequency splitting''
commonly used in quantum field theory.
Section~\ref{secdesitter} treats the Dirac operator in
the de Sitter space-time. In this case, the positive and negative
spectral subspaces of the operator~$\Sig_m$ give a non-trivial interpolation between the
spaces of positive and negative frequency as experienced by observers
at asymptotic times~$t \rightarrow \pm \infty$.
In all these examples, the main task is to prove the mass oscillation
properties. Establishing the weak mass oscillation property will always be an intermediate step
for proving the strong mass oscillation property. 

We finally remark that~\eqref{ppsiMin} and~\eqref{aim} can also be written
with a Dirac distribution as
\beq \label{delta}
\bra \psi_m | \phi_{m'} \ket = \delta(m-m')\: ( \psi_m | \Sig_m \phi_m )_m \:.
\eeq
Such ``$\delta$-normalizations in the mass parameter'' are commonly used in
the perturbative treatment (see~\cite{sea, grotz} and~\cite[\S2.1]{PFP} or more recently~\cite{norm}).
The mass oscillation property makes it possible to give such normalizations
a rigorous meaning in the non-perturbative treatment.

\section{Preliminaries} \label{secprelim}
As in~\cite{finite}, we let~$(\scrM, g)$ be a smooth, globally hyperbolic Lorentzian spin
manifold of dimension~$k \geq 2$.
For the signature of the metric we use the convention~$(+ ,-, \ldots, -)$.
We denote the corresponding spinor bundle by~$S\scrM$. Its fibres~$S_x\scrM$ are endowed
with an inner product~$\Sl .|. \Sr_x$ of signature~$(n,n)$
with~$n=2^{[k/2]-1}$ (where~$[.]$ is the Gau{\ss} bracket; for details see~\cite{baum, lawson+michelsohn}),
which we refer to as the spin scalar product.
Clifford multiplication is described by a mapping~$\gamma$
which satisfies the anti-commutation relations,
\[ \gamma \::\: T_x\scrM \rightarrow \Lin(S_x\scrM) \qquad
\text{with} \qquad \gamma(u) \,\gamma(v) + \gamma(v) \,\gamma(u) = 2 \, g(u,v)\,\1_{S_x(\scrM)} \:. \]
We again write Clifford multiplication in components with the Dirac matrices~$\gamma^j$
and use the short notation with the Feynman dagger, $\gamma(u) \equiv u^j \gamma_j \equiv \uslsh$.
The metric connections on the tangent bundle and the spinor bundle are denoted by~$\nabla$.
The sections of the spinor bundle are also referred to as wave functions.
We denote the smooth sections of the spinor bundle by~$C^\infty(\scrM, S\scrM)$.
Similarly, $C^\infty_0(\scrM, S\scrM)$ denotes the smooth sections with compact support.
On the wave functions, one has the Lorentz invariant inner product
\begin{gather}
\bra .|. \ket \::\: C^\infty(\scrM, S\scrM) \times C^\infty_0(\scrM, S\scrM) \rightarrow \C \:, \notag \\
\bra \psi|\phi \ket = \int_\scrM \Sl \psi | \phi \Sr_x \: d\mu_\scrM\:.  \label{stip} 
\end{gather}

The Dirac operator~$\Dir$ is defined by
\[ \Dir := i \gamma^j \nabla_j + \B \::\: C^\infty(\scrM, S\scrM) \rightarrow C^\infty(\scrM, S\scrM)\:, \]
where~$\B \in \Lin(S_x)$ (the ``external potential'') can be any smooth and symmetric multiplication operator.
For a given real parameter~$m \in \R$ (the ``mass''), the Dirac equation reads
\beq \label{Dirac}
(\Dir - m) \,\psi_m = 0 \:.
\eeq
For clarity, we always denote solutions of the Dirac equation by a subscript~$m$.
We mainly consider solutions in the class~$\Cisc(\scrM, S\scrM)$ of smooth sections
with spatially compact support. On such solutions, one has the scalar product
\beq \label{print}
(\psi_m | \phi_m)_m = 2 \pi \int_\scrN \Sl \psi_m | \nuslsh \phi_m \Sr_x\: d\mu_\scrN(x) \:,
\eeq
where~$\scrN$ denotes any Cauchy surface and~$\nu$ its future-directed normal
(due to current conservation, the scalar product is
in fact independent of the choice of~$\scrN$; for details see~\cite[Section~2]{finite}).
Forming the completion gives the Hilbert space~$(\H_m, (.|.)_m)$.

The {\em{retarded}} and {\em{advanced Green's operators}}~$s_m^\wedge$ and~$s_m^\vee$ are
mappings (for details see for example~\cite{baer+ginoux})
\[ s_m^\wedge, s_m^\vee \::\: C^\infty_0(\scrM, S\scrM) \rightarrow \Cisc(\scrM, S\scrM)\:. \]
Taking their difference gives the so-called causal fundamental solution~$k_m$,
\beq \label{kmdef}
k_m := \frac{1}{2 \pi i} \left( s_m^\vee - s_m^\wedge \right) \::\: C^\infty_0(\scrM, S\scrM) \rightarrow \Cisc(\scrM, S\scrM)
\cap \H_m \:.
\eeq
These operators can be represented as integral operators with distributional kernels; for example,
\[ (k_m \phi)(x) = \int_\scrM k_m(x,y)\, \phi(y)\: d\mu_\scrM(y)\:. \]
The operator~$k_m$ is useful for two reasons. First, it can be used to
construct a solution of the Cauchy problem:
\begin{Prp} \label{prp21} Let~$\scrN$ be any Cauchy surface. Then the solution of the Cauchy problem
\[ (\Dir - m) \,\psi_m = 0 \:,\qquad \psi|_{\scrN} = \psi_\scrN \in C^\infty(\scrN, S\scrM) \]
has the representation
\[ \psi_m(x) = 2 \pi \int_\scrN k_m(x,y)\, \nuslsh\, \psi_\scrN(y)\: d\mu_\scrN(y)\:. \]
\end{Prp} \noindent
Second, the operator~$k_m$ can be regarded as the signature operator of the
inner product~\eqref{stip} when expressed in terms of the scalar product~\eqref{print}:
\begin{Prp} \label{prpdual}
For any~$\psi_m \in \H_m$ and~$\phi \in C^\infty_0(\scrM, S\scrM)$,
\[ (\psi_m \,|\, k_m \phi)_m = \bra \psi_m | \phi \ket \:. \]
\end{Prp} \noindent
Proposition~\ref{prp21} is stated and proved in~\cite[Section~2]{finite}.
For the proof of Proposition~\ref{prpdual} we refer to~\cite[Proposition~2.2]{dimock3} or~\cite[Section~3.1]{finite}.

\section{The Weak Mass Oscillation Property} \label{secwMOP}
\subsection{Basic Definitions} \label{secdefwmop}
In a space-time of infinite life time, the space-time inner product~$\bra \psi_m | \phi_m \ket$ of two
solutions~$\psi_m, \phi_m \in \H_m$ is in general ill-defined, because the time integral in~\eqref{stip}
may diverge. In order to avoid this difficulty, we shall consider 
families of solutions with a variable mass parameter. The so-called mass oscillation property
will make sense of the space-time integral in~\eqref{stip} after integrating over the mass parameter.

More precisely, we consider the mass parameter in a bounded open interval, $m \in I := (m_L, m_R)$.
For a given Cauchy surface~$\scrN$, we consider a function~$\psi_\scrN(x,m) \in S_x\scrM$
with~$x \in \scrN$ and~$m \in I$. We assume that this wave function is smooth and has
compact support in both variables, $\psi_\scrN \in C^\infty_0(\scrN \times I, S\scrM)$.
For every~$m \in I$, we let~$\psi(.,m)$ be the solution of the Cauchy problem for initial data~$\psi_\scrN(.,m)$,
\beq \label{cauchy2}
(\Dir - m) \,\psi(x,m) = 0 \:,\qquad \psi(x,m) = \psi_\scrN(x,m)  \;\; \forall\: x \in \scrN \:.
\eeq
Since the solution of the Cauchy problem is smooth and depends smoothly on parameters,
we know that~$\psi \in C^\infty(\scrM \times I, S\scrM)$.
Moreover, due to finite propagation speed, $\psi(.,m)$ has spatially compact support.
Finally, the solution is clearly compactly supported in the mass parameter~$m$.
We summarize these properties by writing
\beq \label{CscmH}
\psi \in \Cisco(\scrM \times I, S\scrM) \:,
\eeq
where~$\Cisco(\scrM \times I, S\scrM)$ denotes the smooth wave functions with spatially compact support which
are also compactly supported in~$I$. 
We often denote the dependence on~$m$ by a subscript, $\psi_m(x) := \psi(x,m)$.
Then for any fixed~$m$, we can take the scalar product~\eqref{print}. 
On families of solutions~$\psi, \phi \in \Cisco(\scrM \times I, S\scrM)$ of~\eqref{cauchy2},
we introduce a scalar product by integrating over the mass parameter,
\beq \label{spm}
( \psi | \phi) := \int_I (\psi_m | \phi_m)_m \: dm
\eeq
(where~$dm$ is the Lebesgue measure). Forming the completion gives the
Hilbert space~$(\H, (.|.))$. It consists of measurable functions~$\psi(x,m)$
such that for almost all~$m \in I$, the function $\psi(.,m)$ is a weak solution of the Dirac
equation which is square integrable over any Cauchy surface.
Moreover, this spatial integral is integrable over~$m \in I$, so that
the scalar product~\eqref{spm} is well-defined. We denote the norm on~$\H$
by~$\| . \|$.

For the applications, it is useful to introduce
a subspace of the solutions of the form~\eqref{CscmH}:

\begin{Def} \label{defHinf}
We let~$\H^\infty \subset \Cisco(\scrM \times I, S\scrM) \cap \H$ be a subspace
of the smooth solutions with the following properties:
\begin{itemize}
\item[(i)] $\H^\infty$ is invariant under multiplication by smooth functions in the mass parameter,
\[ \eta(m)\, \psi(x,m) \in \H^\infty \qquad \forall\: \psi \in \H^\infty,\;\eta \in C^\infty(I) \:. \]
\item[(ii)] The set~$\H^\infty_m := \{ \psi(.,m) \,|\, \psi \in \H^\infty\}$ is a dense subspace of~$\H_m$, i.e.
\[ \overline{\H^\infty_m}^{(.|.)_m} = \H_m \qquad \forall \:m \in I \:. \]
\end{itemize}
We refer to~$\H^\infty$ as the {\bf{domain}} for the mass oscillation property.
\end{Def} \noindent
The simplest choice is to set~$\H^\infty = \Cisco(\scrM \times I, S\scrM) \cap \H$, but in some applications
it is preferable to choose~$\H^\infty$ as a proper subspace of~$\Cisco(\scrM \times I, S\scrM) \cap \H$.

Our motivation for considering a variable mass parameter is that
integrating over the mass parameter should improve the decay properties
of the wave function for large times (similar as explained in the introduction
in the vacuum Minkowski space).
This decay for large times should also make it possible to integrate the Dirac operator
in the inner product~\eqref{stip} by parts without boundary terms,
\[ \bra \Dir \psi | \phi \ket = \bra \psi | \Dir \phi \ket \:, \]
implying that the solutions for different mass parameters
should be orthogonal with respect to this inner product.
Instead of acting with the Dirac operator, it is technically easier to
work with the operator of multiplication by~$m$, which we denote by
\[ T \::\: \H \rightarrow \H \:,\qquad (T \psi)_m = m \,\psi_m \:. \]
In view of property~(ii) in Definition~\ref{defHinf}, this operator leaves~$\H^\infty$ invariant,
\[ T|_{\H^\infty} \::\: \H^\infty \rightarrow \H^\infty \:. \]
Moreover, $T$ is a symmetric operator, and it is bounded because the interval~$I$ is,
\beq \label{Tsymm}
T^* = T \in \Lin(\H) \:.
\eeq
Finally, integrating over~$m$ gives the operation
\[ \p \::\: \H^\infty \rightarrow \Cisc(\scrM, S\scrM)\:,\qquad
\p \psi = \int_I \psi_m\: dm \:. \]
The next definition should be regarded as specifying
the minimal requirements needed for the construction of the fermionic projector
(stronger assumptions which give rise to additional properties of the fermionic projector
will be considered in Section~\ref{secsMOP} below).

\begin{Def} \label{defWMass Oscillation Property}
The Dirac operator~$\Dir$ on the globally hyperbolic manifold~$(\scrM, g)$ has the {\bf{weak mass
oscillation property}} in the interval~$I \subset \R$ with domain~$\H^\infty$ (see Definition~\ref{defHinf})
if the following conditions hold:
\begin{itemize}
\item[(a)] For every~$\psi, \phi \in \H^\infty$, the
function~$\Sl \p \phi | \p \psi \Sr$ is integrable on~$\scrM$.
Moreover, for any~$\psi \in \H^\infty$ there is a constant~$c(\psi)$ such that
\beq \label{mbound}
|\bra \p \psi | \p \phi \ket| \leq c\, \|\phi\| \qquad \forall\: \phi \in \H^\infty \:.
\eeq
\item[(b)] For all~$\psi, \phi \in \H^\infty$,
\beq \label{mortho}
\bra \p T \psi | \p \phi \ket = \bra \p \psi | \p T \phi \ket \:.
\eeq
\end{itemize}
\end{Def} \noindent
Clearly, in a given space-time one must verify if the assumptions in this definition
are satisfied. Before explaining in various examples how this can be done 
(see Sections~\ref{secultra} and~\ref{secdesitter}), we now proceed by working out
the consequence of the weak mass oscillation property abstractly.

\subsection{A Self-Adjoint Extension of~$\Sig^2$}
In view of the inequality~\eqref{mbound}, every~$\psi \in \H^\infty$
gives rise to a bounded linear functional on~$\H^\infty$.
By continuity, this linear functional can be uniquely extended to~$\H$.
The Riesz representation theorem allows us to represent
this linear functional by a vector~$u \in \H$, i.e.
\[ (u | \phi) = \bra \p \psi | \p \phi \ket \qquad \forall\: \phi \in \H\:. \]
Varying~$\psi$, we obtain the linear mapping
\[ \Sig \::\: \H^\infty \rightarrow \H \:,\qquad
(\Sig \psi | \phi) = \bra \p \psi | \p \phi \ket \quad \forall\: \phi \in \H\:. \]
This operator is symmetric because
\[  (\Sig \psi | \phi) = \bra \p \psi | \p \phi \ket = (\psi | \Sig \phi) \qquad \forall\: \phi, \psi \in \H^\infty\:. \]
Moreover, \eqref{mortho} implies that the operators~$\Sig$ and~$T$ commute,
\beq \label{STcomm}
\Sig \, T = T\, \Sig \::\: \H^\infty \rightarrow \H \:.
\eeq

For the construction of the fermionic projector we need a spectral calculus for the operator~$\Sig$.
Therefore, we would like to construct a self-adjoint extension of the operator~$\Sig$.
A general method for constructing self-adjoint extensions of symmetric operators
is provided by the Friedrichs extension (see for example~\cite[\S33.3]{lax}).
Since this method only applies to semi-bounded operators, we are led to working with
the operator~$\Sig^2$. We thus introduce the scalar product
\[ \la \psi | \phi \ra_{\Sig^2} = ( \psi | \phi ) + ( \Sig \psi | \Sig \phi) \::\: \H^\infty \times
\H^\infty \rightarrow \C\:. \]
Clearly, the corresponding norm is bounded from below by the norm~$\|.\|$.
Thus, forming the completion gives a subspace of~$\H$,
\beq \label{H0def}
\H_{\Sig^2} := \overline{\H^\infty}^{\la .|. \ra_{\Sig^2}} \subset \H \:.
\eeq

\begin{Prp} Introducing the operator~$\Sig^2$ with domain of definition~$\D(\Sig^2)$ by
\begin{gather*}
\D \big(\Sig^2 \big) = \big\{ u \in \H_{\Sig^2} \quad {\text{such that}} \quad
\big| \la u | \phi \ra_{\Sig^2} \big| \leq c(u)\, \|\phi\| \;\; \forall\: \phi \in \H_{\Sig^2} \big\} \\
\Sig^2 \::\: \D(\Sig^2) \subset \H \rightarrow \H \:,\qquad
( \Sig^2 \psi | \phi ) = \la \psi | \phi \ra_{\Sig^2} - ( \psi | \phi ) \;\;
\forall\: \phi \in \H_{\Sig^2} \:,
\end{gather*}
this operator is self-adjoint. The operator~$T$ maps~$\D(\Sig^2)$ to itself and commutes with~$\Sig^2$,
\beq \label{S2Tcomm}
\Sig^2 \,T = T\, \Sig^2 \::\: \D(\Sig^2) \rightarrow \H \:.
\eeq
\end{Prp}
\Proof The self-adjointness of~$\Sig^2$ follows exactly as in the standard construction
of the Friedrichs extension (see for example~\cite[Theorem~33.3.4]{lax}
for the operator~$L:=\Sig^2+1$).

Let us show that~$T(\D(\Sig^2)) \subset \D(\Sig^2)$. Thus let~$u \in \D(\Sig^2)$.
Then~$u \in \H_{\Sig^2}$, so that by definition~\eqref{H0def}
there is a series~$u_n \in \H^\infty$
which converges to~$u$ in the topology given by~$\la .|. \ra_{\Sig^2}$.
Next, for any~$\phi \in \Cisco(\scrM \times I, S\scrM)$, we have the inequality
\[  \la T \phi | T \phi \ra_{\Sig^2} = ( T \phi | T \phi ) + ( \Sig T \phi | \Sig T \phi)  
\overset{\eqref{STcomm}}{=} ( T \phi | \phi ) + ( T\Sig \phi | \Sig T \phi) 
\leq \|T\|_{\H}^2\: \la \phi | \phi \ra_{\Sig^2} \:, \]
showing that the operator~$T$ is also bounded on~$\H_{\Sig^2}$.
As a consequence, the series~$T u_n$ converges in~$\H_{\Sig^2}$ to~$T u$.
Moreover, it follows from~\eqref{Tsymm} and~\eqref{STcomm} that
\beq \label{S2Tprod}
\la T u_n | \phi \ra_{\Sig^2} =
( T u_n | \phi ) + ( \Sig T u_n | \Sig \phi) = ( u_n | T \phi ) + ( \Sig u_n | \Sig T \phi) =
\la u_n | T \phi \ra_{\Sig^2} \:.
\eeq
Taking the limit~$n \rightarrow \infty$, it follows that
\[ \big| \la T u | \phi \ra_{\Sig^2} \big| \leq c(u)\: \| T \phi \| \leq c(u)\: \|T\|\: \|\phi\|\:. \]
We conclude that~$T u \in \D(\Sig^2)$.

To prove~\eqref{S2Tcomm}, we first evaluate the operator product on~$u_n$. Then we know
from~\eqref{S2Tprod} and~\eqref{Tsymm} that~$\Sig^2 T u_n = T \Sig^2 u_n$.
Taking the limit~$n \rightarrow \infty$ gives the result.
\QED

The property~\eqref{S2Tcomm} together with the fact that~$T$ is bounded
guarantees that the resolvent of~$\Sig^2$ commutes with~$T$. More specifically,
\[ \big[ (\Sig^2-i)^{-1}, T \big] = -(\Sig^2-i)^{-1}\, \big[\Sig^2, T \big] \, (\Sig-i)^{-1} \:. \]
The operators~$(\Sig^2-i)^{-1}$ and~$T$ are both normal and bounded and commute with
each other. The spectral theorem for bounded commuting normal operators 
(see for example~\cite[Sections~18 and~31.6]{lax}, also cf.~\cite[Section~VIII.5]{reed+simon})
implies that there is a spectral measure~$E$ on~$\sigma(\Sig^2) \times I$ such that
\beq \label{dEdef}
\big( \Sig^2 \big)^p\: T^q = \int_{\sigma(\Sig^2) \times I} \rho^p\, m^q\: dE_{\rho, m}  \qquad \forall\: p,q \in \N\:.
\eeq

\subsection{The Fermionic Projector as an Operator-Valued Measure}
Acting with the operator~$k_m$ as defined in~\eqref{kmdef} for
each~$m$ separately gives the operator
\[ k \::\: C^\infty_0(\scrM \times I, S\scrM) \rightarrow \H\:,\quad
(k \psi)_m = k_m \psi_m \:. \]
This makes it possible to introduce the fermionic projector~$P_\pm$ as an operator-valued measure
on~$I$. Namely, for any~$f \in C^0(I)$ we define
\beq \label{Ppmdef}
\begin{split}
\int_I f(m)\: dP_\pm(m) &= \frac{1}{2} \int_{\sigma(\Sig^2) \times I} f(m)
\left(\rho^{\frac{1}{2}} \pm \Sig \right) \rho^{-\frac{1}{2}} \:dE_{\rho, m} \, k \\
&\;:\; C^\infty_0(\scrM \times I, S\scrM) \rightarrow \H \:.
\end{split}
\eeq

The next proposition explains the normalization of the fermionic projector.
This normalization can be understood as the spatial normalization, expressed in a
functional calculus form (for the spatial normalization see~\cite[Section~2.3]{finite}
or the elementary discussion in~\cite[Section~2]{norm}).
\begin{Prp} {\bf{(normalization)}} Assume that the operator~$\Sig$ commutes with the spectral measure~$E$
in the sense that\footnote{This condition was omitted in the print version of the paper.
However, as pointed out to us by Albert Much, it is not satisfied in general and needs to be
verified in the applications.}%
\[ \Sig \,E_U = E_U\, \Sig \quad \text{on~$\H^\infty$} \qquad \text{for all Borel sets~$U \subset \sigma(\Sig^2) \times I$}\:. \]
Then for any~$s,s' \in \{\pm 1\}$ and all~$f,g \in C^0(I)$ and~$\psi, \phi \in C^\infty_0(\scrM\times I, S\scrM)$,
\begin{align*}
\Big( &\int_I f(m)\: dP_s(m)\, \psi \:\Big|\: \int_I g(m')\: dP_{s'}(m')\, \phi \Big)
= \delta_{s s'} \int_I \overline{f(m)} \,g(m) \:\bra \psi_m \,|\, (dP_s(m)\, \phi)_m \ket \: .
\end{align*}
\end{Prp}
\Proof Using the continuous functional calculus, we obtain
\begin{align*}
\Big( &\int_I f(m)\: dP_\pm(m)\, \psi \:\Big|\: \int_I g(m')\: dP_\pm(m')\, \phi \Big) \\
&= \frac{1}{4} \int_{\sigma(\Sig^2) \times I} \overline{f(m)}\, g(m) \:\rho^{-1} 
\Big(  k(\psi) \:\Big|\: \big(\rho^{\frac{1}{2}} \pm \Sig \big)^2\, dE_{\rho, m}\, k(\phi) \Big) \\
&\overset{(*)}{=} \frac{1}{2} \int_{\sigma(\Sig^2) \times I} \overline{f(m)}\, g(m) \:\rho^{-\frac{1}{2}} 
\Big(  k(\psi) \:\Big|\: (\rho^{\frac{1}{2}} \pm \Sig)\, dE_{\rho, m}\, k(\phi) \Big) \\
&= \Big( k(\psi) \:\Big|\: \int_I \overline{f(m)} \,g(m)\: dP_s(m)\, \phi \Big)
= \int_I \overline{f(m)} \,g(m) \:\bra \psi_m \,|\, (dP_s(m)\, \phi)_m \ket \:,
\end{align*}
where in~($*$) we multiplied out~$(\rho^{\frac{1}{2}} \pm \Sig)^2$ and used that~$\Sig^2=\rho$.
In the last step we applied~\eqref{spm} and Proposition~\ref{prpdual}.
This gives the result in the case~$s=s'$. The calculation for~$s \neq s'$ is similar,
but in~($*$) we get zero.
\QED

The following proposition, which is an immediate consequence of the continuous functional
calculus, explains in which sense our construction is independent of the choice of the interval~$I$.
\begin{Prp} {\bf{(independence of the choice of~$I$)}} \label{prpind}
Suppose that we have two mass intervals
\[ \check{I}=(\check{m}_L, \check{m}_R) \;\subset\; I=(m_L, m_R) \:. \]
We denote all the objects constructed in~$\check{I}$ with an additional check
and let~$\check{\iota}$ and~$\check{\pi}$ be the natural injection and projection operators,
\begin{align*}
\check{\iota} &\::\: \check{\H} \rightarrow \H \:, \qquad
\big(\check{\iota}(\psi)\big)(x,m) = \left\{ \begin{array}{cl} \psi(x,m) & \text{if~$m \in \check{I}$} \\
0 & \text{otherwise}\:. \end{array} \right. \\
\check{\pi} &\::\: \H \rightarrow \check{\H} \:, \qquad
\check{\pi}(\psi) = \psi|_{\scrM \times \check{I}}\:.
\end{align*}
Then
\begin{align*}
\int_{\check{I}} f(m)\: d\check{P}_\pm(m) &= \check{\pi} \int_I f(m)\: dP_\pm(m) \; \check{\iota}
&\hspace*{-1cm}& \forall\: f \in C^0(I) \\
\int_I f(m)\: dP_\pm(m) &= \check{\iota} \int_{\check{I}} f(m)\: d\check{P}_\pm(m) \;\check{\pi}  
&\hspace*{-1cm}& \forall\: f \in C^0_0(\check{I})\:.
\end{align*}
\end{Prp}

\section{The Strong Mass Oscillation Property} \label{secsMOP}

\subsection{Definition and General Structural Results}
\begin{Def} \label{defsMOP}
The Dirac operator~$\Dir$ on the globally hyperbolic manifold~$(\scrM, g)$ has the {\bf{strong mass oscillation
property}} in the interval~$I=(m_L, m_R)$ with domain~$\H^\infty$ (see Definition~\ref{defHinf}), if there is
a constant~$c>0$ such that
\beq \label{mbound2}
|\bra \p \psi | \p \phi \ket| \leq c \int_I \, \|\phi_m\|_m\, \|\psi_m\|_m\: dm
\qquad \forall\: \psi, \phi \in \H^\infty\:.
\eeq
\end{Def}

\begin{Thm} \label{thmsMOP}
The following statements are equivalent:
\begin{itemize}
\item[(i)] The strong mass oscillation property holds.
\item[(ii)] There is a constant~$c>0$ such that for all~$\psi, \phi \in \H^\infty$,
the following two relations hold:
\begin{align}
|\bra \p \psi | \p \phi \ket| &\leq c\, \|\psi\|\, \|\phi\| \label{mb1} \\
\bra \p T \psi | \p \phi \ket &= \bra \p \psi | \p T \phi \ket\:. \label{mb2}
\end{align}
\item[(iii)] There is a family of linear operators~$\Sig_m \in \Lin(\H_m)$ which are uniformly bounded,
\[ \sup_{m \in I} \|\Sig_m\| < \infty\:, \]
such that
\beq \label{Smdef}
\bra \p \psi | \p \phi \ket = \int_I (\psi_m \,|\, \Sig_m \,\phi_m)_m\: dm \qquad
\forall\: \psi, \phi \in \H^\infty\:.
\eeq
\end{itemize}
\end{Thm}
\Proof The implication (iii)$\Rightarrow$(i) follows immediately from the estimate
\[ |\bra \p \psi | \p \phi \ket| \leq \int_I \big| (\psi_m | \Sig_m \phi_m)_m \big|\: dm
\leq \sup_{m \in I} \|\Sig_m\| \int_I \|\psi_m\|_m \:\|\phi\|_m\: dm \:. \]

In order to prove the implication~(i)$\Rightarrow$(ii), we first apply the Schwarz inequality
to~\eqref{mbound2} to obtain
\begin{align*}
|\bra \p \psi | \p \phi \ket| &\leq c \int_I \, \|\phi_m\|_m\, \|\psi_m\|_m\: dm \\
&\leq c \:\Big( \int_I \|\phi_m\|_m^2\: dm \Big)^\frac{1}{2}
\Big( \int_I \|\psi_m\|_m^2\: dm \Big)^\frac{1}{2} = c\: \|\phi\|\, \|\psi\| \:,
\end{align*}
proving~\eqref{mb1}. Next, for given~$N \in \N$ we subdivide the interval~$I=(m_L, m_R)$
by choosing the intermediate points
\[ m_\ell = \frac{\ell}{N}\: (m_R-m_L) + m_L\:,\qquad \ell=0, \ldots, N\:. \]
Moreover, we choose non-negative test functions~$\eta_1, \ldots, \eta_N \in C^\infty_0(\R)$ which
form a partition of unity and are supported in small subintervals, meaning that
\beq \label{part}
\sum_{\ell=1}^N \eta_\ell \big|_I = 1|_I \qquad \text{and} \qquad \supp \eta_\ell \subset (m_{\ell-2}, m_{\ell+1}) \:,
\eeq
where we set~$m_{-1}=m_L-1$ and~$m_{N+1}=m_R+1$. For any smooth function~$\eta \in C^\infty_0(\R)$ we
define the operator~$\eta(T) \in \Lin(\H) \::\: \H^\infty \rightarrow \H^\infty$ by
\[ \big( \eta(T) \psi \big)_m = \eta(m)\: \psi_m \:. \]
Then by linearity,
\begin{align*}
&\bra \p T \psi | \p \phi \ket - \bra \p \psi | \p T \phi \ket \\
&= \sum_{\ell, \ell'=1}^N \Big( \bra \p \,T \,\eta_{\ell}(T)\, \psi \,|\, \p \,\eta_{\ell'}(T)\, \phi \ket
- \bra \p \,\eta_{\ell}(T)\, \psi \,|\, \p \,T \,\eta_{\ell'}(T)\, \phi \ket \Big) \\
&= \sum_{\ell, \ell'=1}^N 
\Big( \bra \p \,\big(T-m_\ell \big) \,\eta_{\ell}(T)\, \psi \,|\, \p \,\eta_{\ell'}(T)\, \phi \ket
- \bra \p \,\eta_{\ell}(T)\, \psi \,|\, \p \,\big( T-m_\ell \big) \,\eta_{\ell'}(T)\, \phi \ket \big) \:.
\end{align*}
Taking the absolute value and applying~\eqref{mbound2}, we obtain
\[ \big| \bra \p T \psi | \p \phi \ket - \bra \p \psi | \p T \phi \ket \big|
\leq c \sum_{\ell, \ell'=1}^N
\int_I | m-m_\ell | \:\eta_\ell(m)\, \eta_{\ell'}(m)\: \|\phi_m\|_m\, \|\psi_m\|_m\: dm \:. \]
In view of the second property in~\eqref{part}, we only get a contribution if~$|\ell-\ell'| \leq 1$.
Moreover, we know that~$| m-m_\ell | \leq 2 \,|I|/N$ on the support of~$\eta_\ell$. Thus
\begin{align*}
\big| \bra \p T \psi | \p \phi \ket - \bra \p \psi | \p T \phi \ket \big|
&\leq \frac{6 c\,|I|}{N} \sum_{\ell=1}^N
\int_I \eta_\ell(m) \, \|\phi_m\|_m\, \|\psi_m\|_m\: dm \\
&= \frac{6 c \,|I|}{N}
\int_I \|\phi_m\|_m\, \|\psi_m\|_m\: dm \:.
\end{align*}
Since~$N$ is arbitrary, we obtain~\eqref{mb2}.

It remains to prove the implication~(ii)$\Rightarrow$(iii). Combining~\eqref{mb1}
with the Fr{\'e}chet-Riesz theorem, there is a bounded operator~$\Sig \in \Lin(\H)$ with
\[ \bra \p \psi | \p \phi \ket = (\psi | \Sig \phi) \qquad \forall\: \psi, \phi \in \H^\infty\:. \]
The relation~\eqref{mb2} implies that the operators~$\Sig$ and~$T$ commute.
Moreover, these two operators are obviously symmetric and thus self-adjoint.
Hence the spectral theorem for commuting self-adjoint operators implies that there is a spectral measure~$F$
on~$\sigma(\Sig) \times I$ such that
\beq \label{dFdef}
\Sig^p\, T^q = \int_{\sigma(\Sig) \times I} \nu^p \,m^q\, dF_{\nu, m} \qquad \forall\: p,q \in \N\:.
\eeq
For given~$\psi, \phi \in \H^\infty$, we introduce the Borel measure~$\mu_{\psi, \phi}$
on~$I$ by
\beq \label{mudef}
\mu_{\psi, \phi}(\Omega) = \int_{\sigma(\Sig) \times \Omega} \nu\: d(\psi | F_{\nu, m} \phi) \:.
\eeq
Then~$\mu_{\psi, \phi}(I) = (\psi | \Sig \phi)$ and
\[ \mu_{\psi, \phi}(\Omega) = \int_{\sigma(\Sig) \times I} \nu\: d \big( \chi_\Omega(T)\, \psi \,\big|\, 
F_{\nu, m}\, \chi_\Omega(T)\, \phi \big) = (\chi_\Omega(T)\, \psi \,|\, \Sig \,\chi_\Omega(T)\, \phi) \:. \]
Since the operator~$\Sig$ is bounded, we conclude that
\begin{align}
|\mu_{\psi, \phi}(\Omega)| &\leq c \,\|\chi_\Omega(T)\, \psi\|\, \|\chi_\Omega(T)\, \phi\| 
\overset{\eqref{spm}}{=} c \left( \int_\Omega \|\psi\|_m^2 \:dm \; \int_\Omega  \|\psi\|_{m'}^2 \:dm' \right)^\frac{1}{2}
\notag \\
&\leq c\, |\Omega| \:\Big( \sup_{m \in \Omega} \|\psi_m\|_m \Big) \Big( \sup_{m' \in \Omega} \|\phi_{m'} \|_{m'} \Big)\:.
\label{RNes}
\end{align}
This shows that the measure~$\mu$ is absolutely continuous with respect to the Lebesgue measure.
The Radon-Nikodym theorem (see~\cite[Theorem~6.9]{rudin} or~\cite[\S VI.31]{halmosmt})
implies that there is a unique function~$f_{\psi, \phi} \in L^1(I, dm)$ such that
\beq \label{fppdef}
\mu_{\psi, \phi}(\Omega) = \int_\Omega f_{\psi, \phi}(m)\: dm\:.
\eeq
Moreover, the estimate~\eqref{RNes} gives the pointwise bound
\[ | f_{\psi, \phi}(m) | \leq c\,\|\psi_m\|_m \: \|\phi_m \|_m \:. \]
Using this inequality, we can apply the Fr{\'e}chet-Riesz theorem to obtain a unique
operator~$\Sig_m \in \Lin(\H_m)$ such that
\beq \label{Smdef2}
f_{\psi, \phi}(m) = (\psi_m | \Sig_m \phi_m)_m \qquad \text{and} \qquad
\|\Sig_m\| \leq c\:.
\eeq
Combining the above results, for any~$\psi, \phi \in \H^\infty$ we obtain
\begin{align*}
\bra \p \psi | \p \phi \ket &= (\psi | \Sig \phi)
=\int_{\sigma(\Sig) \times I} \nu\: d( \psi \,|\, F_{\nu, m}\, \phi ) \\
&= \int_I d\mu_{\psi, \phi} = \int_I f_{\psi, \phi}(m)\: dm
= \int_I (\psi_m | \Sig_m \phi_m)_m\: dm \:.
\end{align*}
This concludes the proof.
\QED

Comparing the statement of Theorem~\ref{thmsMOP}~(ii) with Definition~\ref{defWMass Oscillation Property},
we immediately obtain the following result.
\begin{Corollary} The strong mass oscillation property implies the weak mass oscillation property.
\end{Corollary}

We next show uniqueness as well as the independence of the choice of the interval~$I$.
\begin{Prp} {\bf{(uniqueness of~$\Sig_m$)}} \label{prpunique}
The family~$(\Sig_m)_{m \in I}$ in the statement of Theorem~\ref{thmsMOP}
can be chosen such that for all~$\psi, \phi \in \H^\infty$, the expectation
value~$ f_{\psi, \phi}(m) := (\psi_m | \Sig_m \phi_m)_m$ is continuous in~$m$,
\beq \label{flip}
f_{\psi, \phi} \in C^0_0(I) \:.
\eeq
The family~$(\Sig_m)_{m \in I}$ with the properties~\eqref{Smdef} and~\eqref{flip} is unique.
Moreover, choosing two intervals~$\check{I}$ and~$I$ with~$m \in \check{I} \subset I$
and~$0 \not \in \overline{I}$, 
and denoting all the objects constructed in~$\check{I}$ with an additional check,
we have
\beq \label{ScS}
\check\Sig_m = \Sig_m \:.
\eeq
\end{Prp}
\Proof Let us show that the function~$f_{\psi, \phi}$ is continuous.
To this end, we choose a function~$\eta \in C^\infty_0(I)$. Then for any~$\varepsilon>0$
which is so small that~$B_\varepsilon(\supp \eta) \subset I$, we obtain
\begin{align*}
\int_I & \Big( f_{\psi, \phi}(m+\varepsilon) - f_{\psi, \phi}(m) \Big) \:\eta(m)\: dm
= \int_I f_{\psi, \phi}(m) \:\Big( \eta(m-\varepsilon) - \eta(m) \Big)\: dm \\
&\overset{(\ast)}{=} \bra \int_I \Big( \eta(m-\varepsilon) - \eta(m) \Big) \psi_m\: dm \:|\: \p \phi \ket
= \bra \int_I \eta(m) \: \Big( \psi_{m+\varepsilon} - \psi_m \Big)\: dm \:|\: \p \phi \ket \:,
\end{align*}
where in~($\ast$) we used~\eqref{dFdef} and~\eqref{mudef}. Applying~\eqref{mb1}, we obtain
\[ \left| \int_I \Big( f_{\psi, \phi}(m+\varepsilon) - f_{\psi, \phi}(m) \Big) \:\eta(m)\: dm \right|
\leq c \, \|\psi_{+\varepsilon} - \psi\|\: \|\phi\|\: \sup_I |\eta| \:, \]
where the vector~$\psi_{+\varepsilon} \in \H^\infty$ is defined by~$(\psi_{+\varepsilon})_m := \psi_{m+\varepsilon}$.
Since~$\lim_{\varepsilon \searrow 0} \|\psi_{+\varepsilon} - \psi\| = 0$ and~$\eta$ is
arbitrary, we conclude that~$f_{\psi, \phi}$ is continuous~\eqref{flip}.
This continuity is important because it implies that the function~$f_{\psi, \phi}$ is uniquely defined
pointwise (whereas in~\eqref{fppdef} this function could be modified arbitrarily on sets of measure zero).

In order to prove~\eqref{ScS}, we first note that the spectral
measures~$dE_{\rho, m}$ and~$dF_{\nu, m}$ (cf.~\eqref{dEdef} and~\eqref{dFdef})
are related to each other by
\[ dE_{\rho, m} = dF_{\sqrt{\rho}, m} + dF_{-\sqrt{\rho}, m} \:. \]
A direct computation yields that the definitions~\eqref{Ppmdef} and~\eqref{Ppmdef2} agree
if the strong mass oscillation property holds (see also~\eqref{mudef}, \eqref{fppdef} and~\eqref{Smdef2}).
The relation~\eqref{ScS} then follows from Proposition~\ref{prpind}.
\QED
We remark that by considering higher difference quotients and taking the limit~$\varepsilon \searrow 0$,
one could even prove that~$f_{\psi, \phi} \in C^\infty_0(I)$ is smooth, but this is not of relevance here.

\subsection{Construction of the Fermionic Projector}
Theorem~\ref{thmsMOP} and Proposition~\ref{prpunique}
are very useful because for every~$m \in I$ they provide a unique
operator~$\Sig_m \in \Lin(\H_m)$, referred to as the {\bf{fermionic signature operator}}
corresponding to the mass~$m$. This makes it possible to proceed 
with methods similar to~\cite{finite}.
From Definition~\ref{Smdef}, the operator~$\Sig_m$ is obviously symmetric.
Thus the spectral theorem gives rise to the spectral decomposition
\[ \Sig_m = \int_{\sigma(\Sig_m)} \nu\: dE_\nu \:, \]
where~$E_\nu$ is the spectral measure (see for example~\cite{reed+simon}).
The spectral measure gives rise to the spectral calculus
\[ f(\Sig_m) = \int_{\sigma(\Sig_m)} f(\nu)\: dE_\nu \:, \]
where~$f$ is a bounded Borel function.

\begin{Def} \label{ferm_proj_Def}
Assume that the Dirac operator~$\Dir$ on~$(\scrM,g)$ 
satisfies the strong mass oscillation property (see Definition~\ref{defsMOP}).
We define the operators~$P_\pm \::\: C^\infty_0(\scrM, S\scrM) \rightarrow \H_m$ by
\beq \label{Ppmdef2}
P_+ = \chi_{[0, \infty)}(\Sig_m)\, k_m \qquad \text{and} \qquad P_- = -\chi_{(-\infty, 0)}(\Sig_m)\, k_m
\eeq
(where~$\chi$ denotes the characteristic function).
The {\bf{fermionic projector}} $P$ is defined by~$P=P_-$.
\end{Def}

\begin{Prp} \label{prpPpm} 
For all~$\phi, \psi \in C^\infty_0(\scrM, S\scrM)$, the operators~$P_\pm$ are symmetric,
\[ \bra P_\pm \phi \,|\, \psi \ket = \bra \phi \,|\, P_\pm \psi \ket \:. \]
Moreover, the image of~$P_\pm$ is the positive respectively negative
spectral subspace of~$\Sig_m$, i.e.
\beq \label{dense}
\overline{P_+(C^\infty_0(\scrM, S\scrM))} = E_{(0, \infty)}(\H_m) \:,\qquad
\overline{P_-(C^\infty_0(\scrM, S\scrM))} = E_{(-\infty, 0)}(\H_m) \:.
\eeq
\end{Prp}
\Proof According to Proposition~\ref{prpdual},
\begin{align*}
\bra P_- \phi \,|\, \psi \ket = (P_- \phi \,|\, k_m \psi )_m
&= - \big( \chi_{(-\infty, 0)}(\Sig_m)\, k_m \phi \,\big|\, k_m \psi \big)_m \\
&= - \big( k_m\,\phi \,\big|\, \chi_{(-\infty, 0)}(\Sig_m)\, k_m \psi \big)_m
= \bra \phi \,|\, P_- \psi \ket \:.
\end{align*}
The proof for~$P_+$ is similar.
The relations~\eqref{dense} follow immediately from the fact
that~$k_m(C^\infty_0(\scrM, S\scrM))$ is dense in~$\H_m$.
\QED

\subsection{Representation as a Distribution and Normalization}
Similar as in~\cite[Theorem~3.12]{finite}, the fermionic projector 
can be represented by a two-point distribution on~$\scrM$. As usual,
we denote the space of test functions (with the Fr{\'e}chet topology)
by~$\D$ and define the space of distributions~$\D'$ as its dual space.
\begin{Thm} Assume that the strong mass oscillation property holds. Then
there is a unique distribution~${\mathcal{P}} \in \D'(\scrM \times \scrM)$ such that 
for all~$\phi, \psi \in C^\infty_0(\scrM, S\scrM)$,
\[ \bra \phi | P \psi \ket = {\mathcal{P}}(\phi \otimes \psi) \:. \]
\end{Thm}
\Proof According to Proposition~\ref{prpdual} and Definition~\ref{ferm_proj_Def},
\[ \bra \phi | P \psi \ket = (k_m \phi \,|\, P \psi) = - (k_m \phi \,|\, \chi_{(-\infty, 0)}(\Sig_m) \,k_m \psi) \:. \]
Since the norm of the operator~$\chi_{(-\infty, 0)}(\Sig_m)$ is bounded by one, we conclude that
\[ |\bra \phi | P \psi \ket| \leq \| k_m \phi\| \:\|k_m \psi\| 
= ( \bra \phi | k_m \phi \ket \: \bra \psi | k_m \psi \ket )^\frac{1}{2} \:, \]
where in the last step we again applied Proposition~\ref{prpdual}.
As~$k_m \in \D'(\scrM \times \scrM)$, the right side is continuous on~$\D(\scrM \times \scrM)$.
We conclude that also~$\bra \phi | P \psi \ket$ is continuous on~$\D(\scrM \times \scrM)$.
The result now follows from the Schwartz kernel theorem (see~\cite[Theorem~5.2.1]{hormanderI},
keeping in mind that this theorem applies just as well to bundle-valued distributions on a manifold
simply by working with the components in local coordinates and a local trivialization).
\QED
Exactly as explained in~\cite[Section~3.5]{finite}, it is convenient to use the standard notation with an integral kernel~$P(x,y)$,
\begin{align*}
\bra \phi | P \psi \ket &= \iint_{\scrM \times \scrM} \Sl \phi(x) \,|\, P(x,y) \,\psi(y) \Sr_x \: d\mu_\scrM(x)\: d\mu_\scrM(y) \\
(P \psi)(x) &= \int_{\scrM} P(x,y) \,\psi(y) \: d\mu_\scrM(y)
\end{align*}
(where~$P(.,.)$ coincides with the distribution~${\mathcal{P}}$ above).
In view of Proposition~\ref{prpPpm}, we know that the last integral is not only a distribution,
but a function which is square integrable over every Cauchy surface.
Moreover, the symmetry of~$P$ shown in Proposition~\ref{prpPpm} implies that
\[ P(x,y)^* = P(y,x) \:, \]
where the star denotes the adjoint with respect to the spin scalar product.
Finally, exactly as shown in~\cite[Proposition~3.13]{finite},
the spatial normalization property of Proposition~\ref{prpspatnorm} makes it possible
to obtain a representation of the fermionic projector in terms of one-particle states.
To this end, one chooses an orthonormal basis~$(\psi_j)_{j \in \N}$
of the subspace~$\chi_{(-\infty, 0)}(\Sig_m) \subset \H_m$. Then
\[ P(x,y) = -\sum_{j=1}^\infty |\psi_j(x) \Sr \Sl \psi_j(y)| \]
with convergence in~$\D'(\scrM \times \scrM)$.

We now specify the normalization of the fermionic projector.
We introduce an operator~$\Pi$ by
\beq \label{Pidef}
\Pi \::\: \H_m \rightarrow \H_m \:, \qquad
(\Pi \,\psi_m)(x) = -2 \pi \int_\scrN P(x,y)\, \nuslsh\, (\psi_m)|_\scrN(y)\: d\mu_\scrN(y)  \:,
\eeq
where~$\scrN$ is any Cauchy surface.
\begin{Prp} {\bf{(spatial normalization)}} \label{prpspatnorm}
The operator~$\Pi$ is a projection operator on~$\H_m$.
\end{Prp}
\Proof According to Proposition~\ref{prp21}, the spatial integral in~\eqref{Pidef}
can be combined with the factor~$k_m$ in~\eqref{Ppmdef2} to give the solution of the
corresponding Cauchy problem. Thus
\[ \Pi \::\: \H_m \rightarrow \H_m \:, \qquad
(\Pi \,\psi_m)(x) = \chi_{(-\infty, 0)}(\Sig_m)\, \psi_m \:, \]
showing that~$\Pi$ is a projection operator.
\QED

Instead of the spatial normalization, one could also consider the mass normalization
(for details on the different normalization methods see~\cite{norm}). To this end, one needs
to consider families of fermionic projectors~$P_m$ indexed by the mass parameter. Then
for all~$\phi, \psi \in C^\infty_0(\scrM, S\scrM)$, we can use~\eqref{Smdef} and Proposition~\ref{prpdual}
to obtain
\begin{align*}
\bra \p (P_m \phi) \,|\, \p (P_{m'} \psi) \ket &= \int_I (P_m \phi \,|\, \Sig_m P_m \psi)_m\: dm
= \int_I (k_m \phi \,|\, \Sig_m \chi_{(-\infty, 0)}(\Sig_m)\, k_m \psi)_m\: dm \\
&= \int_I \bra \phi \,|\, \Sig_m \chi_{(-\infty, 0)}(\Sig_m)\, k_m \psi \ket\: dm = -\bra \phi \,|\, \p(\Sig_m P_m \psi) \ket \:,
\end{align*}
which can be written in a compact formal notation similar to~\eqref{delta} as
\[ P_m \,P_{m'} = \delta(m-m')\: (-\Sig_m) \, P_m \:. \]
Due to the factor~$(-\Sig_m)$ on the right, in general the fermionic projector does {\em{not}} satisfy the
mass normalization condition.
The mass normalization condition could be arranged by modifying the definition~\eqref{Ppmdef2} to
\[ \Sig_m^{-1}\, \chi_{(-\infty, 0)}(\Sig_m)\, k_m \:. \]
Here we prefer to work with the spatial normalization. For a detailed discussion of the
different normalization methods we refer to~\cite[Section~2]{norm}.

We finally remark that corresponding causal fermion systems can be constructed
exactly as in~\cite[Section~4]{finite} by introducing regularization
operators~$({\mathfrak{R}}_\varepsilon)_{\varepsilon>0}$, computing the local correlation
operators~$F^\varepsilon(x)$ and defining the universal measure by~$d\rho=F^\varepsilon_* d\mu_\scrM$.

\section{Example: Ultrastatic Space-Times} \label{secultra}
In this section we prove that the strong mass oscillation property holds for 
the Dirac operator in complete ultrastatic space-times, even if an arbitrary static magnetic field is present.
Thus we let~$(\scrM,g)$ be a $k$-dimensional complete space-time
which is ultrastatic in the sense that
it is the product~$\scrM= \R \times \scrN$ with a metric of the form
\[ ds^2 = dt^2 -g_\scrN \:, \]
where $g_\scrN$ is a Riemannian metric on~$\scrN$.
The completeness of~$\scrM$ implies that also~$\scrN$ is complete.
Moreover, we assume that~$\scrN$ is spin.
Let $\Dir_\scrN$ denote the intrinsic Dirac operator on~$\scrN$.
In order to introduce the magnetic field, we let~$A$ be a smooth vector field on~$\scrN$
(the ``vector potential'') and set
\beq \label{Dira}
\Dir_A = \Dir_\scrN + \slashed{A} \:,
\eeq
where the slash again denotes Clifford multiplication.
Using standard elliptic theory (see~\cite[Proposition~8.2.7]{taylor2} and~\cite{chernoff}),
the operator~$\Dir_A$ with domain~$C^\infty_0(\scrN, S\scrN)$ is
essentially self-adjoint on the Hilbert space~$L^2(\scrN, S\scrN)$. Thus its closure, which we again denote
by~$\Dir_A$, is a self-adjoint operator with domain~$\D(\Dir_A)$. The spectral theorem yields
\beq \label{ultrast_Dirac-op_spectraldecomp}
\Dir_A = \int_{\sigma(\Dir_A)} \lambda\: dF_\lambda \:,
\eeq
where $dF_\lambda$ denotes the spectral-measure of $\Dir_A$.

The Dirac operator in the ultrastatic space-time~$(\scrM,g)$ in the presence of the magnetic field~$A$ 
can be written in block matrix notation as
\beq \label{Dirultra}
\Dir = \begin{pmatrix} i \partial_t & -\Dir_A \\ \Dir_A & -i \partial_t \end{pmatrix} .
\eeq
Since the Dirac operator is time independent, we can separate the time dependence with a plane
wave ansatz,
\[ \psi(t,x) = e^{-i \omega t}\: \chi(x)\:. \]
The sign of~$\omega$ gives a natural decomposition of the solution space into two
subspaces. This is often referred to as ``frequency splitting,'' and the subspaces
are called the solutions of positive and negative energy, respectively.

This is the main result of this section.
\begin{Thm} \label{thmultrastrong} 
On any interval~$I = (m_L,m_R)$ with~$m_L,m_R > 0$,
the Dirac operator~\eqref{Dirultra} has the strong mass oscillation property
with domain
\beq \label{Hinfchoice}
\H^\infty := \Cisco(\scrM \times I, S\scrM) \cap \H \:.
\eeq
The operators~$\Sig_m$ in the representation~\eqref{Smdef} all have the spectrum~$\{\pm 1\}$.
The eigen\-spa\-ces corresponding to the eigenvalues~$\pm 1$
coincide with the solutions of positive and negative frequency, respectively.
\end{Thm} \noindent
We remark that the reason why the spectral decomposition of~$\Sig_m$ gives the frequency splitting
can already be understood in the perturbative treatment as explained in~\cite[Section~5]{sea}.
As a corollary, the above theorem clearly yields the strong mass oscillation property for the
Dirac operator in the Minkowski vacuum.

We now begin with preparations for the proof, which will be completed at the end of Section~\ref{secmstrong}.
The space-time inner product~\eqref{stip} and the scalar product~\eqref{print} take the form
\begin{align}
\bra \phi | \psi \ket &= \int_{-\infty}^\infty dt \int_\scrN \Sl \psi | \phi \Sr_{(t,x)}\: d\mu_\scrN(x) 
= \int_{-\infty}^\infty dt\: \Big\la \psi \Big| \begin{pmatrix} 1 & 0 \\ 0 & -1 \end{pmatrix} \phi \Big\ra_{L^2(\scrN, S\scrN)^2} \\
( \phi | \psi )_m &= \int_\scrN \Sl \psi | \begin{pmatrix} 1 & 0 \\ 0 & -1 \end{pmatrix} \phi \Sr_{(t,x)}\: d\mu_\scrN(x) 
= 2 \pi \,\la \psi(t) | \phi(t) \ra_{L^2(\scrN, S\scrN)^2} \label{mprod}
\end{align}
(where in the last line~$t$ is arbitrary due to current conservation).
In the following constructions, we will also work with the last scalar product without
requiring that~$\phi$ and~$\psi$ are solutions of the Dirac equation. In this case,
the scalar product will depend on time, and we denote it by
\[ ( \phi | \psi )_t = 2 \pi \,\la \psi(t) | \phi(t) \ra_{L^2(\scrN, S\scrN)^2} \:. \]
We usually write the Dirac equation in the Hamiltonian form as
\[ i \partial_t \psi = H \psi \qquad \text{with} \qquad
H=\left(\begin{array}{cc} 0 & \Dir_A \cr \Dir_A & 0 \end{array}  \right) + m \left(\begin{array}{cc} \1 & 0 \cr 0 & -\1 \end{array}  \right) . \]
Substituting the spectral decomposition \eqref{ultrast_Dirac-op_spectraldecomp}, we get
\[ H = \int_{\sigma(\Dir_A)} \begin{pmatrix} m & \lambda \cr \lambda & -m \end{pmatrix}\, dF_\lambda \:. \]
In order to bring the dynamics into a more explicit form, we diagonalize the $2 \times 2$-matrix,
\[ \begin{pmatrix} m & \lambda \cr \lambda & -m \end{pmatrix} = \omega\, \Pi_+
-  \omega\, \Pi_- \:, \]
where we set
\beq \label{wdef}
\omega = \sqrt{\lambda^2 + m^2} \:.
\eeq
The matrices and~$\Pi_\pm$ are orthogonal projections, i.e.
\[ \Pi_s\: \Pi_{s'} = \delta_{ss'}\: \Pi_s \qquad \forall\: s,s' \in \{\pm\}\:. \]
A short computation shows that
\beq \label{ultrast_proj}
\Pi_\pm = \Pi_\pm(\lambda,m) = \frac{\1}{2} \pm \frac{1}{2 \omega} \begin{pmatrix}
m & \lambda \cr \lambda & -m  \end{pmatrix} .
\eeq
Applying the functional calculus, the solution of the Dirac equation of mass~$m$
with initial data $\psi_m|_{t=0} = \psi_m(0) \in C^\infty_0(\scrN, S\scrM)$ can be written as
\beq \label{psim}
\psi_m(t) = e^{-itH(m)} \psi_m(0) = \int_{\sigma(\Dir_A)} U_m^t(\lambda) \:dF_\lambda \: \psi_m(0) \:,
\eeq
where~$U_m^t$ is the unitary $2 \times 2$-matrix
\beq \label{Utdef}
U_m^t(\lambda) = e^{-it \omega(\lambda, m)} \:\Pi_+(\lambda, m)
+ e^{it\omega(\lambda, m)} \:\Pi_-(\lambda, m) \:.
\eeq

\subsection{The Weak Mass Oscillation Property using Mass Derivatives} \label{secmweak}
In preparation for the strong mass oscillation property, we shall now prove the
weak mass oscillation property. Let~$\psi \in \H^\infty$ as defined in~\eqref{Hinfchoice}. Then
\beq \label{ppsi}
(\p \psi)(t) = \int_I dm \int_{\sigma(\Dir_A)} U^t_m(\lambda) \:dF_\lambda \: \psi_m(0) \:.
\eeq
For estimates of such expressions, it is helpful to observe that~$U^t_m(\lambda)$
is a $2 \times 2$-matrix which commutes with the spectral measure~$dF_\lambda$.
In particular, the matrix entries of the inner integral in~\eqref{ppsi} can be written as
\beq \label{gdef}
g(m) := \int_{\sigma(\Dir_A)} f(\lambda, m) \:dF_\lambda \: \psi(m) \;\in\; L^2(\scrN, S\scrN)
\eeq
with~$f \in C^\infty(I \times \R)$ and~$\psi \in C^\infty_0(\scrN \times I, S\scrN)$
(where we use the notation~$\psi(m) = \psi_m(.) \in C^\infty(\scrN, S\scrN)$).
In the next lemma it is shown that this function
is differentiable and that we may interchange the differentiation with the integral.
Since this is a somewhat subtle point, we give the proof in detail.
\begin{Lemma} \label{lemmasmooth}
Let~$\psi \in C^\infty_0(\scrN \times I, S\scrN)$ be a smooth family of wave functions on~$\scrN$.
Moreover, let~$f \in C^\infty(I \times \R)$ be a smooth function such that~$f$ and all
its mass derivatives are polynomially bounded, i.e.\ for all~$p \in \N$ there is~$\ell \in \N$ 
and a constant~$c>0$ such that
\beq \label{fassum}
|\partial_m^p f(\lambda, m)| \leq c \: \big( 1+\lambda^{2 \ell} \big) \qquad \forall\:
\lambda \in \R, \:m \in I \:.
\eeq
Then the function~$g$ defined by~\eqref{gdef} satisfies the bound
\beq \label{gbound}
\big\| g(m) \big\|_{L^2(\scrN, S\scrN)} \leq c \,\big\| \big(1+\Dir_A^{2 \ell} \big) \psi \big\|_{L^2(\scrN, S\scrN)} \:.
\eeq
Moreover, the function~$g$ is smooth in~$m$ and
\beq \label{gprime}
g^{(p)}(m) = \int_{\sigma(\Dir_A)} dF_\lambda \: \partial_m^p \Big( f(\lambda, m) \: \psi(m) \Big) \:.
\eeq
\end{Lemma}
\Proof For the proof of the bound~\eqref{gbound}, we may omit the mass dependence. Then
the spectral calculus yields
\begin{align*}
\Big\| & \int_{\sigma(\Dir_A)} f(\lambda) \:dF_\lambda \: \psi \Big\|^2_{L^2(\scrN, S\scrN)}
= \int_{\sigma(\Dir_A)} |f(\lambda)|^2 \:d \la \psi | F_\lambda \psi \ra_{L^2(\scrN, S\scrN)} \\
&\overset{\eqref{fassum}}{\leq} c \int_{\sigma(\Dir_A)} \big(1+\lambda^{2 \ell} \big)^2 \:d \la \psi | 
F_\lambda \psi \ra_{L^2(\scrN, S\scrN)}
= c \,\big\| \big(1+\Dir_A^{2 \ell} \big) \psi \big\|^2_{L^2(\scrN, S\scrN)} \:.
\end{align*}

In order to prove that~$g$ is differentiable, we consider the difference quotient and subtract
the expected derivative,
\[ \phi_\varepsilon := \frac{g(m+\varepsilon)-g(m)}{\varepsilon}
-\int_{\sigma(\Dir_A)} dF_\lambda \: \partial_m \big( f(\lambda, m) \: \psi(m) \big) . \]
By rearranging the terms, we obtain
\begin{align}
\phi_\varepsilon &= \int_{\sigma(\Dir_A)} dF_\lambda \,\bigg[ \frac{f(\lambda, m+\varepsilon) \:
\psi(m+\varepsilon) -f(\lambda, m) \: \psi(m)}{\varepsilon} - \partial_m \Big( f(\lambda, m) \: \psi(m) \Big) \bigg]
\notag \\
&= \int_{\sigma(\Dir_A)} dF_\lambda \,\bigg[ \Big( \frac{f(\lambda, m+\varepsilon) -f(\lambda, m)}{\varepsilon}
-\partial_m f(\lambda, m) \Big)\: \psi(m) \Big) \label{t1} \\
&\hspace*{2.6cm} + f(\lambda, m+\varepsilon) \:\Big\{ \frac{\psi(m+\varepsilon) - \psi(m)}{\varepsilon} -
\partial_m \psi(m) \Big\} \label{t2} \\
&\hspace*{2.6cm} + \Big( f(\lambda, m+\varepsilon) - f(\lambda, m) \Big) \:\partial_m \psi(m) \bigg] \:. \label{t3}
\end{align}
The contribution~\eqref{t2} can be estimated immediately with the help of~\eqref{gbound}
(with the function~$\psi(m)$ in~\eqref{gdef} replaced by the expression in the curly brackets in~\eqref{t2}). We thus obtain
\[ \| \eqref{t2} \|_{L^2(\scrN, S\scrN)} \leq 
c \: \Big\| \big(1+\Dir_A^{2 \ell} \big) \Big( \frac{\psi(m+\varepsilon) - \psi(m)}{\varepsilon} -
\partial_m \psi(m) \Big) \Big\|_{L^2(\scrN, S\scrN)}\, , \]
and this converges to zero as~$\varepsilon \searrow 0$ because~$\psi$ is smooth
and has compact support. The term~\eqref{t3}, on the other hand, is estimated by
decomposing the $\lambda$-integral into the integrals over the regions~$[-L,L]$
and~$\R \setminus [-L,L]$ and estimating similar as in the proof of~\eqref{gbound},
\begin{align}
\Big\| & \int_{-L}^L dF_\lambda \, \Big( f(\lambda, m+\varepsilon) - f(\lambda, m) \Big) \:\partial_m \psi(m) \Big\|_{L^2(\scrN, S\scrN)} \notag \\
& \leq \|\partial_m \psi(m)\|_{L^2(\scrN, S\scrN)} \sup_{(\lambda, m) \in [-L,L] \times I}
\big| f(\lambda, m+\varepsilon) - f(\lambda, m) \big| \:. \label{t31} \\
\intertext{Moreover, using again~\eqref{fassum},}
\Big\| & \int_{\R \setminus [-L,L]} dF_\lambda \, \Big( f(\lambda, m+\varepsilon) - f(\lambda, m) \Big)
\:\partial_m \psi(m) \Big\|_{L^2(\scrN, S\scrN)}^2 \notag \\
&\leq 4c^2 \int_{\R \setminus [-L,L]} \big(1+\lambda^{2 \ell} \big)^2 \:d \big\la \partial_m \psi(m), F_\lambda 
\partial_m \psi(m) \big\ra_{L^2(\scrN, S\scrN)} \notag \\
&\leq \frac{4c^2}{L^4} \int_{\R \setminus [-L,L]} \big(1+\lambda^{2 \ell+2} \big)^2 \:d \big\la \partial_m \psi(m), F_\lambda \partial_m \psi(m) \big\ra_{L^2(\scrN, S\scrN)} \notag \\
&= \frac{4c^2}{L^4} \: \big\| \big(1+\Dir_A^{2 \ell+2} \big) \psi \big\|_{L^2(\scrN, S\scrN)}^2 \:. \label{t32}
\end{align}
The term~\eqref{t32} can be made arbitrarily small by choosing~$L$ sufficiently large.
The term~\eqref{t31}, on the other hand, tends to zero as~$\varepsilon \searrow 0$ for any fixed~$L$ due to
the locally uniform convergence of~$f(\lambda, m+\varepsilon)$ to~$f(\lambda, m)$
(note that~$f$ is smooth in view of~\eqref{Utdef} and~\eqref{ultrast_proj}).
This shows that~\eqref{t3} tends to zero as~$\varepsilon \searrow 0$.
Finally, the contribution~\eqref{t1} can be estimated just as~\eqref{t3} by
considering the regions~$[-L,L]$ and~$\R \setminus [-L,L]$ separately.

We conclude that in the limit~$\varepsilon \searrow 0$,
the vectors~$\phi_\varepsilon$ converge to zero in~$L^2(\scrN, S\scrN)$.
This shows~\eqref{gprime} in the case~$p=1$. The relation for general~$p$
follows immediately by induction.
\QED

\begin{Lemma} \label{lemmaint2}
The time evolution operator in the vacuum has the representation
\beq \label{Udecay}
t^2 \,U_m^t(\lambda) = \frac{\partial^2}{\partial m^2} A_m^t(\lambda) + 
\frac{\partial}{\partial m} B_m^t(\lambda) + C_m^t(\lambda)
\eeq
with matrices~$A_m^t$, $B_m^t$ and~$C_m^t$ which are bounded
uniformly in time by
\[ \|A_m^t(\lambda)\| + \|B_m^t(\lambda)\| + \|C_m^t(\lambda)\| 
\leq c \:\big( 1 + \lambda^2 \big) \qquad \forall\: m \in I \]
with a constant~$c$ which may depend on the choice of the interval~$I$
(here~$\| . \|$ denotes any norm on the $2 \times 2$-matrices, and we
again assume that~$I = (m_L,m_R)$ with~$m_L,m_R > 0$).
\end{Lemma}
\Proof We can generate factors of~$t$ by differentiating the exponentials in~\eqref{Utdef}
with respect to~$\omega$. With the help of~\eqref{wdef}, we can then rewrite the
$\omega$-derivatives as $m$-derivatives. We thus obtain
\[ t^2 \, e^{\pm i \omega t} = -\frac{\partial^2}{\partial \omega^2} e^{\pm i \omega t}
= -\frac{\omega}{m} \frac{\partial}{\partial m} \left( \frac{\omega}{m} \frac{\partial}{\partial m}
e^{\pm i \omega t} \right) \:. \]
A straightforward computation in which one uses the product rule inductively
gives the result.
\QED

\begin{Lemma} \label{lemmappsi}
For any~$\psi \in \H^\infty$, there is a constant~$C=C(\psi)$ such that
\[ \big\| (\p \psi)|_t \big\|_{t} \leq \frac{C}{1+t^2} \:. \]
\end{Lemma}
\Proof Using that the operators~$U^t_m$ are unitary, we immediately obtain
\[ \big\|(\p \psi)|_t \big\|_{t} \leq \int_I dm \:\|\psi_m\|_m \:. \]
In order to prove time decay, we apply the identity~\eqref{Udecay} to~\eqref{ppsi}.
Then Lemma~\ref{lemmasmooth} allows us to integrate by parts,
\begin{align*}
t^2 (\p \psi)|_t &= \int_\scrM dm \int_{\sigma(\Dir_A)}  dF_\lambda
\left( \frac{\partial^2}{\partial m^2} A_m^t + 
\frac{\partial}{\partial m} B_m^t + C_m^t \right) \psi_m(0) \\
&= \int_\scrM dm \int_{\sigma(\Dir_A)} dF_\lambda\:
\Big( A_m^t(\lambda) \:\partial_m^2 \psi_m(0) - B_m^t(\lambda) \:\partial_m \psi_m(0)
+ C_m^t(\lambda) \:\psi_m(0) \Big) .
\end{align*}
Now can use the estimate of Lemma~\ref{lemmaint2} together with~\eqref{gbound} to obtain
\begin{align}
t^2 \, \big\|(\p \psi)|_t \big\|_{t} &\leq c \int_\scrM dm 
\sum_{a=0,2} \bigg\| \int_{\sigma(\Dir_A)} (1+\lambda^2) \:dF_\lambda\:
\partial_m^a \psi_m(0) \bigg\|_{t} \nonumber \\
&= c \int_\scrM dm 
\sum_{a=0,2} \Big\| (1+\Dir_A^2)\, \partial_m^a \psi_m(0) \Big\|_{t} \:, \label{psider}
\end{align}
where in the last step we used the spectral calculus.
\QED

\begin{Prp} \label{prpwMOP}
On any interval~$I = (m_L,m_R)$ with~$m_L,m_R > 0$,
the Dirac operator~\eqref{Dirultra} has the weak mass oscillation property
with domain~\eqref{Hinfchoice}.
\end{Prp}
\Proof
For every~$\psi, \phi \in \H^\infty$, the Schwarz inequality gives
\[ |\bra \p \psi | \p \phi \ket| = \left| \int_{-\infty}^\infty \big( (\p \psi)|_t \,\big|\, \gamma^0 \,(\p \phi)|_t \big)_t \:dt \right| 
\leq \int_{-\infty}^\infty \big\| (\p \psi)|_t \big\|_t\: \big\| (\p \phi)|_t \big\|_t\: dt \:. \]
Applying Lemma~\ref{lemmappsi} together with the estimate
\begin{align*}
\big\| (\p \phi)|_t \big\|^2_t &= \iint_{I \times I} (\phi_m | \phi_{m'})_t \:dm \,dm' \\
&\leq \frac{1}{2} \iint_{I \times I} \Big( \|\phi_m\|^2 +  \|\phi_{m'}\|^2 \Big)\: dm \,dm' = |I| \,\|\phi\|^2 \:,
\end{align*}
we obtain the inequality~\eqref{mbound} with
\[ c = C\,\sqrt{|I|} \int_{-\infty}^\infty \frac{dt}{1+t^2} < \infty \:. \]
 
The identity~\eqref{mortho} follows by integrating the Dirac operator in space-time by parts,
\beq \label{pintD}
\begin{split}
\bra \p T \psi | \p \phi \ket &= \bra \p \Dir \psi | \p \phi \ket =
\bra \Dir \p \psi | \p \phi \ket = \int_\scrM \Sl \Dir \p \psi | \p \phi \Sr(x)\: d^4x \\
&\overset{(\ast)}{=} \int_\scrM \Sl \p \psi | \Dir \p \phi \Sr(x)\: d^4x = \bra \p \psi | \Dir \p \phi \ket
= \bra \p \psi | \p T \phi \ket \:.
\end{split}
\eeq
In~$(\ast)$ we used that the Dirac operator is formally self-adjoint with respect to~$\bra .|. \ket$.
Moreover, we do not get boundary terms in view of the time decay in Lemma~\ref{lemmappsi}.
\QED

\subsection{The Strong Mass Oscillation Property using a Plancherel Method} \label{secmstrong}
We now give the proof of Theorem~\ref{thmultrastrong}.
Before beginning, we point out that the method of working with mass derivatives
in the previous section gave the inequality~\eqref{mbound} with a constant~$c$
which depended on the derivatives of~$\psi$ (cf.~\eqref{psider}).
For the strong mass oscillation property, however, this constant must depend only
on the $L^2$-norm of~$\psi$ (see~\eqref{mb1}).
For this reason, working with mass derivatives and an integration-by-parts
argument in the mass parameter is not appropriate for proving the strong mass oscillation property.
Instead, we shall use the following Plancherel method.

First, in view of the decay established in Lemma~\ref{lemmappsi},
we know that for any~$\psi, \phi \in \H^\infty$, the function~$\Sl \p \psi | \p \phi \Sr$ is
integrable. Moreover, the time integral can be carried out with the help of Plancherel's theorem,
\begin{align}
\bra \p \psi | \p\phi \ket &= \int_{-\infty}^\infty dt \int_\scrN \Sl \p \psi | \p \phi \Sr_{(t,x)} \:d\mu(x) \notag \\
&= \int_{-\infty}^\infty \frac{d\omega}{2 \pi} \: \Big\la \widehat{\p \psi}(\omega) \Big| 
\begin{pmatrix} 1 & 0 \\ 0 & -1 \end{pmatrix} \widehat{\p \phi}(\omega) \Big\ra_{L^2(\scrN, S\scrN)^2} \:,
\label{ultrast_eqn1}
\end{align}
where
\[ \widehat{\p \psi}(\omega) = \int_{-\infty}^\infty (\p \psi)(t)\: e^{i \omega t}\: dt \:. \]
In order to compute this Fourier transform, we take the representation~\eqref{psim} and~\eqref{Utdef},
integrate over the mass parameter, and rewrite the mass integral as an integral over~$\omega$,
\begin{align*}
&(\p \psi)(t) = \sum_{s=\pm} \int_I dm \int_{\sigma(\Dir_A)} e^{-sit \omega(\lambda, m)} \:\Pi_s(\lambda, m) \:dF_\lambda \: \psi_m(0) \\
&= \sum_{s=\pm} \int_0^\infty \frac{dm}{d\omega} \:d\omega
\int_{\sigma(\Dir_A)} \!\!\!\! \chi_{(m_L^2, m_R^2)}\big(\omega^2-\lambda^2 \big)  \:e^{- s i \omega t}\:
\Pi_s(\lambda, m) \:dF_\lambda \: \psi_m(0) \Big|_{m=\sqrt{\omega^2-\lambda^2}} \:,
\intertext{where the characteristic function~$\chi_{(m_L^2, m_R^2)}\big(\omega^2-\lambda^2 \big)$
vanishes unless~$\sqrt{\omega^2-\lambda^2} \in I$.
Using~\eqref{wdef}, we obtain}
&= \int_{-\infty}^\infty d\omega \:e^{-i \omega t}\: \frac{|\omega|}{m}
\int_{\sigma(\Dir_A)} \!\!\!\! \chi_{(m_L^2, m_R^2)}\big(\omega^2-\lambda^2 \big) \:
\Pi_s(\lambda, m) \:dF_\lambda \: \psi_m(0) \bigg|_{m=\sqrt{\omega^2-\lambda^2},\; s=\text{sign}(\omega)} \:.
\end{align*}
This shows that
\[ \widehat{\p \psi}(\omega) = 2 \pi\: \frac{|\omega|}{m}
\int_{\sigma(\Dir_A)} \chi_{(m_L^2, m_R^2)} \big(\omega^2-\lambda^2 \big)\: 
\Pi_s(\lambda, m) \:dF_\lambda \: \psi_m(0) \bigg|_{m=\sqrt{\omega^2-\lambda^2},\;
s=\text{sign}(\omega)} \:. \]

Using this formula in~\eqref{ultrast_eqn1} and applying the spectral calculus for~$\Dir_A$, we obtain
\begin{align*}
&\bra \p \psi | \p\phi \ket = 2 \pi  \int_{-\infty}^\infty d\omega\: \frac{\omega^2}{m^2}
\int_{\sigma(\Dir_A)} \chi_{(m_L^2, m_R^2)} \big(\omega^2-\lambda^2 \big) \\
&\times
\Big\la \Pi_s(\lambda, m) \:\psi_m(0) \Big| 
\begin{pmatrix} 1 & 0 \\ 0 & -1 \end{pmatrix} dF_\lambda \:\Pi_s(\lambda, m) \:\phi_m(0) \Big\ra_{L^2(\scrN, S\scrN)^2}
\Big|_{m=\sqrt{\omega^2-\lambda^2},\;s=\text{sign}(\omega)}\:.
\end{align*}
A short computation using~\eqref{ultrast_proj} shows that
\[ \Pi_s(\lambda, m) \begin{pmatrix} 1 & 0 \\ 0 & -1 \end{pmatrix} \Pi_s(\lambda, m)
= \frac{m}{s \,|\omega|}\: \Pi_s(\lambda, m) \:. \]
Hence
\begin{align*}
\bra \p \psi | \p\phi \ket &= 2 \pi  \int_{-\infty}^\infty d\omega\: \frac{\omega}{m}
\int_{\sigma(\Dir_A)} \chi_{(m_L^2, m_R^2)} \big(\omega^2-\lambda^2 \big) \\
&\qquad \qquad \times
\big\la \psi_m(0) \big|  dF_\lambda \:\Pi_s(\lambda, m) \:\phi_m(0) \big\ra_{L^2(\scrN, S\scrN)^2}
\Big|_{m=\sqrt{\omega^2-\lambda^2},\;s=\text{sign}(\omega)} \\
&= 2 \pi \int_I dm \sum_{s=\pm} s \int_{\sigma(\Dir_A)}
d\big\la \psi_m(0) \,\big|\, F_\lambda\: \Pi_s(\lambda, m) \:\phi_m(0) \big\ra_{L^2(\scrN, S\scrN)^2} \:,
\end{align*}
where in the last step we transformed back to the integration variable~$m$.
Using~\eqref{mprod}, we obtain the identity
\beq \label{freqsplit}
\bra \p \psi | \p\phi \ket = \int_I dm \int_{\sigma(\Dir_A)}
d \big( \psi_m \,\big|\, F_\lambda\: \big( \Pi_+(\lambda, m) - \Pi_-(\lambda,m) \big) \:\phi_m \big)_m \:.
\eeq
The inequality~\eqref{mbound2} follows immediately by applying the Schwarz inequality
and using that the matrices~$\Pi_\pm$ have norm one.

Finally, comparing~\eqref{freqsplit} with~\eqref{Smdef}, one sees
that the eigenvalues and corresponding eigenspaces of the operator~$\Sig_m$
coincide precisely with those of the matrix~$\Pi_+ - \Pi_-$.
Hence~$\Sig_m$ has the spectrum~$\{\pm 1\}$, and 
in view of~\eqref{Utdef} the eigenspaces are precisely
the subspaces of positive and negative frequency.
This completes the proof of Theorem~\ref{thmultrastrong}.

\section{Example: De Sitter Space-Time} \label{secdesitter}
We consider the de Sitter space-time~$\scrM=\R \times S^3$ with the line element
\beq \label{deSitter}
ds^2 = dt^2 - R(t)^2\: ds^2_{S^3} \qquad \text{and} \qquad R(t) = \cosh t \:,
\eeq
where~$ds^2_{S^3}$ is the line element on the three-dimensional unit sphere.
This is a special case of the Friedmann-Robertson-Walker metric
obtained for a specific choice of the scaling function. The Dirac operator was
computed in~\cite{moritz} (see also~\cite[Appendix~A]{bounce}) to be
\beq
\Dir = i \gamma^0 \left( \partial_t + \frac {3 \dot{R}(t)}{2R(t)} \right) + \frac{1}{R(t)}
\begin{pmatrix} 0 & \Dir_{S^3} \\ -\Dir_{S^3} & 0 \\ \end{pmatrix} , \label{dirac}
\eeq
where~$\Dir_{S^3}$ is the Dirac operator on~$S^3$. The inner products~\eqref{stip}
and~\eqref{print} take the form
\begin{align}
\bra \psi | \phi \ket &= \int_{-\infty}^\infty dt \int_{S^3} \Sl \psi | \phi \Sr(t,x) \:R(t)^3\, d\mu_{S^3}(x) \label{stip2} \\
( \psi_m | \phi_m )_m &= 2 \pi \int_{S^3} \Sl \psi | \gamma^0 \phi \Sr(t,x) \:R(t)^3\,  d\mu_{S^3}(x) \:, \label{print2}
\end{align}
where~$\Sl \psi | \phi \Sr = \psi^\dagger \gamma^0 \phi$ and~$\gamma^0=\diag(1,1,-1,-1)$
(here~$d\mu_{S^3}$ is the normalized volume measure on~$S^3$).

In order to separate the Dirac equation~\eqref{Dirac}, one uses that, being an elliptic operator
on a bounded domain, the Dirac operator on~$S^3$
has a purely discrete spectrum and finite-dimensional eigenspaces.
More specifically, the eigenvalues are (see~\cite{lawson+michelsohn} or the detailed computations in~\cite[Appendix~A]{moritz}),
\[ \sigma(\Dir_{S^3}) = \Big\{ \pm \frac{3}{2}, \,\pm \frac{5}{2}, \,\pm \frac{7}{2}, \ldots \Big\} \]
with corresponding eigenspaces of dimensions
\[ \dim \ker ( \Dir_{S^3} - \lambda) = \lambda^2 - \frac{1}{4}\:. \]
Since the Dirac operator~\eqref{dirac} obviously commutes with~$\Dir_{S^3}$, 
the solution spaces can be decomposed into eigenspaces of~$\Dir_{S^3}$. We use the notation
\[ \H_m = \bigoplus_{\lambda \in \sigma(\Dir_{S^3})} \H_m^{(\lambda)}\:,\qquad
\H = \bigoplus_{\lambda \in \sigma(\Dir_{S^3})} \H^{(\lambda)} \:. \]
We also refer to the eigenspaces of~$\Dir_{S^3}$ as {\em{spatial modes}}.
Next, we choose~$\H^\infty$ as the proper subspace of~$\Cisco(\scrM \times S, S\scrM) \cap \H$
of solutions composed of a {\em{finite number of spatial modes}},
\beq \label{Hinfchoice2}
\H^\infty = \Big\{ \psi \in \Cisco(\scrM \times S, S\scrM) \cap \H \;\Big|\;
\psi \in \bigoplus\nolimits_{|\lambda| \leq \Lambda} \H^{(\lambda)} \text{ with } \Lambda \in \R \Big\}
\eeq
(this choice clearly has all the properties demanded in Definition~\ref{defHinf};
the reason for this choice will be explained after Lemma~\ref{lemma31} below).
This is our main result:
\begin{Thm} \label{thmdeSitter} On any interval~$I=(m_L, m_R)$ with~$m_L, m_R>0$,
the Dirac operator in the de Sitter space-time has the strong
mass oscillation property with domain~\eqref{Hinfchoice2}.
\end{Thm}

The remainder of this section is devoted to the proof of this theorem.
Choosing a normalized eigenspinor~$\phi^{(\lambda)}$ of~$\Dir_{S^3}$ corresponding
to the eigenvalue~$\lambda$, we employ the ansatz
\begin{equation} \label{ansatz}
\psi_m = R(t)^{-\frac{3}{2}}
\begin{pmatrix} u_1(m,t) \:\phi^{(\lambda)}(x) \\
u_2(m,t) \:\phi^{(\lambda)}(x) \end{pmatrix}
\end{equation}
to obtain the coupled system of ordinary differential equations
\beq \label{DiracODE}
i\, \frac{d}{dt} \begin{pmatrix} u_1 \\ u_2 \end{pmatrix}
=  \begin{pmatrix} m & -\lambda/R \\ -\lambda/R & - m \end{pmatrix}
\begin{pmatrix} u_1 \\ u_2 \end{pmatrix}
\eeq
for the complex-valued functions~$u_1$ and~$u_2$.
The inner products~\eqref{stip2} and~\eqref{print2} become
\begin{align}
\bra \psi | \tilde{\psi} \ket &= \int_{-\infty}^\infty \left(\overline{u_1} \tilde{u}_1 - \overline{u_2} \tilde{u}_2 \right) dt  \label{stip3} \\
( \psi_m | \tilde{\psi}_m )_m &= 2 \pi \left( \overline{u_1} \tilde{u}_1 + \overline{u_2} \tilde{u}_2 \right) 
= 2 \pi \la u, \tilde{u} \ra_{\C^2} \:. \label{print3}
\end{align}
Using that the matrix in~\eqref{DiracODE} is Hermitian, one easily
verifies that
\beq \label{currcons}
\frac{d}{dt} \la u, \tilde{u} \ra_{\C^2} = 0\:,
\eeq
showing that the scalar product~\eqref{print3} is indeed time independent.
We refer to a wave function of the form~\eqref{ansatz} as a {\em{single spatial mode}}.

The asymptotics of  solutions of ~\eqref{DiracODE} for large times can be described
with a simple Gr\"onwall-type estimate:
\begin{Lemma} \label{lemma31}
Asymptotically as $t \rightarrow \pm \infty$, every solution of~\eqref{DiracODE} is of the form
\beq u(t) = \begin{pmatrix} e^{-imt} \:f^\pm_1 \\[0.3em]
e^{imt} \:f^\pm_2 \end{pmatrix} + E^\pm(t) \label{eq:3z}
\eeq
with the error term bounded by
\beq
\|E^\pm(t)\| \leq \|f^\pm\| \, \exp \big( 2 \,|\lambda| \,e^{\mp t} \big) \label{eq:3c}
\eeq
(thus~$E^\pm(t)$ decays exponentially as~$t \rightarrow \pm \infty$).
\end{Lemma}
\Proof Substituting into~\eqref{DiracODE} the ansatz
\beq
u(t) = \begin{pmatrix} e^{-imt} \:f_1(t) \\
e^{imt} \:f_2(t) \end{pmatrix} \;, \label{eq:36}
\eeq
we obtain for $f$ the differential equation
\beq \label{feq}
\frac{df}{dt} = -\frac{\lambda}{R} \begin{pmatrix}
0 & e^{2imt} \\ e^{-2imt} & 0 \end{pmatrix}  f \:.
\eeq
Taking the norm, we obtain the differential inequality
\beq \label{eq:36a}
\left\| \frac{df}{dt} \right\| \leq \frac{|\lambda|}{R} \:\|f\| \:.
\eeq

Let us first show that~$f(t)$ has a limit as~$t \rightarrow \pm \infty$.
To this end, we first apply Kato's inequality to~\eqref{eq:36a},
\beq \label{kato}
\frac{d}{dt}\, \|f\| \leq \frac{|\lambda|}{R} \:\|f\| \:.
\eeq
We may assume that our solution is nontrivial, so that~$\|f\| \neq 0$.
Thus we may divide by~$\|f\|$,
\[ \frac{d}{dt} \log \|f\| \leq \frac{|\lambda|}{R} \:. \]
Since the scaling function grows exponentially for large~$t$
(cf.~\eqref{deSitter}), we conclude that~$\|f\|$ is bounded and converges as~$t \rightarrow \pm \infty$.
Using this a-priori bound in~\eqref{eq:36a}, we infer that~$f$ has bounded
variation, implying that~$\lim_{t \rightarrow \pm \infty} f$ exists. We set
\[ f^\pm = \lim_{t \rightarrow \pm \infty} f(t) \:. \]

In order to estimate~$\|f-f^-\|$, we divide~\eqref{kato} by~$\|f\|$ and integrate from~$t_0$ to any~$t>t_0$,
\[ \|f(t)\| \leq \|f(t_0)\| \exp \left( \int_{t_0}^{t} \frac{|\lambda|}{R(\tau)}\: d\tau \right) \:. \]
Substituting this inequality into~\eqref{eq:36a} gives
\[ \left\| \frac{df}{dt} \right\| \leq \frac{|\lambda|}{R(t)}\: \|f(t_0)\| \exp \left( \int_{t_0}^{t} \frac{|\lambda|}{R(\tau)}\:
d\tau \right) 
= \|f(t_0)\| \:\frac{d}{dt} \exp \left( \int_{t_0}^{t} \frac{|\lambda|}{R(\tau)}\:
d\tau \right) \:. \]
Integrating on both sides from~$t_0$ to some~$t>t_0$ gives
\[ \left\| f(t) - f(t_0) \right\| \leq \|f(t_0)\| \:\exp \left( \int_{t_0}^{t} \frac{|\lambda|}{R(\tau)}\:
d\tau \right) \:. \]
Now we take the limit~$t_0 \rightarrow -\infty$ to obtain
\[ \left\| f(t) - f^- \right\| \leq \|f(t_0)\| \:\exp \left( \int_{-\infty}^{t} \frac{|\lambda|}{R(\tau)}\:
d\tau \right) \:. \]
Employing the estimate
\[ R(t) = \cosh t \geq \frac{e^{-t}}{2} \:, \]
we conclude that
\[ \left\| f(t) - f^- \right\| \leq \|f(t_0)\| \: \exp \big( 2 |\lambda| \,e^t \big) \:. \]
Using this estimate in~\eqref{eq:36} and comparing with~\eqref{eq:3z} gives the desired estimate for~$E^-$.

The estimate for~$E^+$ is derived similarly.
\QED

As is typical for a Gr\"onwall estimate, the error bound~\eqref{eq:3c} grows exponentially in~$\lambda$.
In particular, our estimate is not uniform in the spatial modes. It is not clear how to improve this estimate
to for example polynomial growth in~$\lambda$. This is the reason why
with the choice~\eqref{Hinfchoice2} we always restrict attention to a finite number of spatial modes.

\begin{Lemma} \label{lemmafsmooth}
For every single mode~$\psi \in \Cisco(\scrM \times I, S\scrM) \cap \H^{(\lambda)}$,
the corresponding coefficients~$f^\pm$ in~\eqref{eq:3z} are smooth in~$m$,
\[ f^\pm \in C^\infty_0(I, \C^2) \:. \]
\end{Lemma}
\Proof Evaluating~$\psi$ at time~$t=0$, we get a smooth family~$u(m, t=0)$.
Consequently, the function~$f$ is smooth,
\[ f|_{t=0} \in C^\infty_0(I, \C^2) \:. \]
Taking these initial conditions and solving the equation~\eqref{feq}, we get a family of solutions
which clearly depend smoothly on~$m$. Differentiating~\eqref{feq} with respect to the mass
and setting~$f^{(p)} := \partial_m^p f$, we obtain
\[ \frac{df^{(1)}}{dt} = \frac{\lambda}{R} \begin{pmatrix}
0 & e^{2imt} \\ e^{-2imt} & 0 \end{pmatrix}  f^{(1)} +
2 i t \: \frac{\lambda}{R} \begin{pmatrix}
0 & e^{2imt} \\ -e^{-2imt} & 0 \end{pmatrix}  f \]
and thus
\beq \label{f1diff}
\bigg\| \frac{df^{(1)}}{dt} \bigg\| \leq \frac{|\lambda|}{R}\: \|f^{(1)}\| + 2 |t|\, \frac{|\lambda|}{R}\: \|f\| \:.
\eeq
Again applying the Kato inequality, we obtain similar to~\eqref{kato} the differential inequality
\[ \frac{d}{dt} \left( e^{-\int^t \frac{|\lambda|}{R}} \:\|f^{(1)}(t)\| \right) \leq e^{-\int^t \frac{|\lambda|}{R}}\:
2 |t|\, \frac{|\lambda|}{R(t)}\: \|f(t)\| \:. \]
Integrating on both sides and using the exponential growth of~$R(t)$ at infinity, we conclude
that~$\|f^{(1)}(t)\|$ converges as~$t \rightarrow \pm \infty$. Using this fact in~\eqref{f1diff},
we infer that also the vector~$f^{(1)}(t)$ converges as~$t \rightarrow \pm \infty$.
The higher derivatives~$f^{(p)}$ can be estimated inductively by differentiating~\eqref{feq}
$p$ times with respect to~$m$, taking the norm, and integrating the resulting differential inequality. \QED

\begin{Lemma} \label{lemmapab} For every single mode~$\psi \in \Cisco(\scrM \times I, S\scrM) \cap \H^{(\lambda)}$,
the corresponding function~$u$ in~\eqref{ansatz} satisfies the inequality
\beq \label{pabes}
\big\| (\p u)(t) \big\| \leq \frac{c(\psi)}{1+t^2} \:.
\eeq
\end{Lemma}
\Proof Integrating the asymptotic expansion~\eqref{eq:3z} over the mass parameter gives
\beq \label{pab}
\p u = \int_I \begin{pmatrix} e^{-imt} \:f^\pm_1 \\[0.3em]
e^{imt} \:f^\pm_2 \end{pmatrix} dm + \int_I E^\pm \: dm\:.
\eeq
The integral over the error term can be estimated by
\beq \label{errbound}
\left\| \int_I E^\pm \: dm \right\| \leq |I|\, \sup_{m \in I} \| E^\pm \| 
\leq c(\psi)\, e^{\mp t}\:,
\eeq
where in the last step we applied~\eqref{eq:3c} and used that~$\sup_m \|f^\pm\|$
is bounded by Lemma~\ref{lemmafsmooth}.
Writing the first summand in~\eqref{pab} as
\[ \int_I \begin{pmatrix} e^{-imt} \:f^\pm_1(t) \\[0.3em]
e^{imt} \:f^\pm_2(t) \end{pmatrix} dm
= -\frac{1}{t^2} \int_I \left[ \frac{d^2}{dm^2} 
\begin{pmatrix} e^{-imt} & 0 \\
0 & e^{imt}\end{pmatrix}
\right] \begin{pmatrix} f^\pm_1 \\[0.3em]
f^\pm_2 \end{pmatrix} dm \:, \]
we can integrate by parts to obtain the estimate
\[ \bigg\| \int_I \begin{pmatrix} e^{-imt} \:f^\pm_1 \\[0.3em]
e^{imt} \:f^\pm_2 \end{pmatrix} dm \bigg\|
\leq \frac{|I|}{t^2}\: \sup_{m \in I} \|\partial_m^2 f^\pm\| \leq \frac{c(\psi)}{t^2}  \:. \]
Combining this estimate with~\eqref{errbound}, we obtain~\eqref{pabes}.
\QED

\begin{Prp}
The Dirac operator in the de Sitter space-time has the weak
mass oscillation property with domain~\eqref{Hinfchoice2}.
\end{Prp}
\Proof Suppose that~$\psi, \phi \in \H^\infty$. Since~$\psi$ and~$\phi$ only involve a finite number of
spatial modes, we may restrict attention to one of them. Moreover, using orthonormality of the
spatial eigenfunctions, we may assume that~$\phi$ and~$\psi$ have the same spatial dependence.
Then the Schwarz inequality yields
\begin{align*}
\| (\p \phi)|_t\| \leq \int_I \| \phi(m) \|_m\: dm \leq \sqrt{|I|}\: \|\phi\| \:.
\end{align*}
Combining this inequality with~\eqref{stip2}, \eqref{print2} and Lemma~\ref{lemmapab}, we obtain
\[ |\bra \p \psi | \p \phi \ket| \leq \sqrt{|I|} \;\|\phi\| \int_{-\infty}^\infty \|(\p u)(t)\|\:dt
\leq \|\phi\| \int_{-\infty}^\infty \frac{c}{1+t^2}\: dt \:, \]
proving~\eqref{mbound}.

The property~\eqref{mortho} follows by integrating the Dirac operator by parts according to~\eqref{pintD},
where in~$(\ast)$ we again use that the Dirac operator is formally self-adjoint with respect to~$\bra .|. \ket$.
Moreover, we do not get boundary terms in view of the time decay in~\eqref{pabes}.
\QED

Our next task is to compute the inner product~$\bra \p \psi | \p \tilde{\psi} \ket$
for two single modes~$\psi, \tilde{\psi} \in \H^{(\lambda)}$ with the same spatial dependence.
We write the result of Lemma~\ref{lemma31} as
\beq \label{psirep}
u(m,t) = \Theta(t) \begin{pmatrix} e^{-imt} \:f^+_1(m) \\[0.3em]
e^{imt} \:f^+_2(m) \end{pmatrix} + \Theta(-t) \begin{pmatrix} e^{-imt} \:f^-_1(m) \\[0.3em]
e^{imt} \:f^-_2(m) \end{pmatrix} + E_m(t) \:,
\eeq
where~$\Theta$ is the Heaviside function,
and the error term decays exponentially as~$t \rightarrow \pm \infty$,
\beq \label{Emes}
\| E_m(t) \| \leq c\: e^{-|t|}\:.
\eeq
For the function~$\tilde{\psi}$ we use the same notation with an additional tilde.

\begin{Lemma} \label{lemmamode}
For any single modes~$\psi, \tilde{\psi} \in \H^{(\lambda)}$ with the same spatial dependence,
\beq \label{modeform}
\bra \p \psi | \p \tilde{\psi} \ket = \pi \sum_{s = \pm} \int_I \left( \overline{f^s_1(m)} \tilde{f}^s_1(m)
- \overline{f^s_2(m)} \tilde{f}^s_2(m) \right) dm \:.
\eeq
\end{Lemma}
\Proof We first explain why the error terms~$E_m(t)$ and~$\tilde{E}_{\tilde{m}}(t)$
do not enter the formula. To this end, we again use the partition of unity~$(\eta_\ell)_{\ell=1,\ldots, N}$
introduced in the proof of Theorem~\ref{thmsMOP} (see~\eqref{part}).
Since we already know that the weak mass oscillation property holds,
we conclude from~\eqref{mortho} inductively that~$\bra \p T^p \psi | \p \phi \ket = \bra \p \psi | \p T^p \phi \ket$
for all~$p$. Using the continuous functional calculus corresponding to the spectral theorem~\eqref{dEdef},
we conclude that
\[ \bra \p\, \eta_\ell(T) \psi \,|\, \p\, \eta_{\ell'}(T) \tilde{\psi} \ket
= \bra \p\, \big(\eta_\ell \,\eta_{\ell'} \big)(T)  \:\psi \,|\, \p \tilde{\psi} \ket \:, \]
which implies by the right side of~\eqref{part} that
\[ \bra \p\, \eta_\ell(T) \psi \,|\, \p\, \eta_{\ell'}(T) \tilde{\psi} \ket = 0 \qquad
\text{unless~$|\ell-\ell'| \leq 1$}\:. \]
Estimating the integrals of the error terms by
\[ \bigg\| \int_I \eta_\ell(m)\, E_m(t) \bigg\| \leq \frac{|I|}{N}\: \sup_m \|E_m(t)\| \]
and using the bound~\eqref{Emes},
the contribution by the error terms tends to zero as~$N \rightarrow \infty$.

It remains to consider the first two summands in~\eqref{psirep}.
Because of the Heaviside functions, we only get the product of~$f_a^+$ with~$\tilde{f}_b^+$
and of~$f_a^-$ with~$\tilde{f}_b^-$. Moreover, the following argument shows why
it suffices to consider the contributions where the lower indices coincide:
For example, the contribution involving~$f_1^+$ and~$\tilde{f}_2^+$ is
\[ A := \int_0^\infty dt \iint_{I \times I} \overline{f_1^+(m)} \,\tilde{f}_2^+(\tilde{m})
\:e^{i (m+\tilde{m}) t} \: dm\: d\tilde{m} \:. \]
Using the distributional equation
\beq \label{ddis}
\int_0^\infty e^{i \omega t} = \pi \delta(\omega) + i\: \frac{\text{PP}}{\omega}
\eeq
(where PP denotes the principal value of the integral), we can use the fact
that~$m+\tilde{m}$ is bounded away from zero to obtain
\[ A = \iint_{I \times I} \frac{\overline{f_1^+(m)} \tilde{f}_2^+(m)}{m+\tilde{m}}\: dm\: d\tilde{m} \:. \]
Again inserting the partition of unity~$(\eta_\ell)_{\ell=1,\ldots, N}$ and taking the limit~$N \rightarrow \infty$
gives zero. The other contributions for~$a \neq b$ are treated similarly.

We conclude that it suffices to take into account the products of the form~$f_a^s$ with~$\tilde{f}_a^s$
with~$a=1,2$ and~$s=\pm 1$. Thus
\begin{align*}
\bra & \p \psi | \p \tilde{\psi} \ket \\
=\:& \int_0^\infty dt \iint_{I \times I}
\Big( e^{i (m-m') t }\: \overline{f_1^+(m)}\: \tilde{f}_1^+(\tilde{m}) -
e^{-i (m-m') t} \: \overline{f_2^+(m)}\: \tilde{f}_2^+(\tilde{m}) \Big) dm\, d\tilde{m} \\
&+ \int_{-\infty}^0 dt \iint_{I \times I}
\Big( e^{i (m-m') t }\: \overline{f_1^-(m)}\: \tilde{f}_1^-(\tilde{m}) -
e^{-i (m-m') t} \: \overline{f_2^-(m)}\: \tilde{f}_2^-(\tilde{m}) \Big) dm\, d\tilde{m} \:,
\end{align*}
and applying~\eqref{ddis} gives
\begin{align}
\bra \p \psi | \p \tilde{\psi} \ket=\:& \pi \int_I \Big( \overline{f_1^+} \tilde{f}_1^+ - \overline{f_2^+} \tilde{f}_2^+ 
+ \overline{f_1^-}\: \tilde{f}_1^- - \overline{f_2^-}\: \tilde{f}_2^- \Big) \big|_m\: dm \\
&+ i \iint_{I \times I} \frac{\text{PP}}{m-\tilde{m}} \Big(
\overline{f_1^+(m)}\: \tilde{f}_1^+(\tilde{m}) + \overline{f_2^+(m)}\: \tilde{f}_2^+(\tilde{m}) \label{c2} \\
&\hspace*{2.75cm} - \overline{f_1^-(m)}\: \tilde{f}_1^-(\tilde{m}) - \overline{f_2^-(m)}\: \tilde{f}_2^-(\tilde{m})
\Big) \:dm\, d\tilde{m} \:. \label{c3}
\end{align}
Using current conservation~\eqref{currcons} together with~\eqref{eq:3z}, 
we may evaluate the scalar product asymptotically as~$t \rightarrow \pm \infty$ to obtain
\[ \sum_{a=1,2} \overline{f_a^+(m)}\: \tilde{f}_a^+(\tilde{m})
= \sum_{a=1,2} \overline{f_a^-(m)}\: \tilde{f}_a^-(\tilde{m}) \:. \]
This implies that the terms~\eqref{c2} and~\eqref{c3} cancel each other, giving the result.
\QED

\Proof[Proof of Theorem~\ref{thmdeSitter}] Suppose that~$\psi, \tilde{\psi} \in \H^\infty$.
Then we decompose them into spatial modes, i.e.
\[ \psi = \sum_{|\lambda| < |\Lambda|} \sum_{k=1}^{K(\lambda)} 
R(t)^{-\frac{3}{2}}
\begin{pmatrix} (u_1)^{(\lambda)}_k(m,t) \:\phi^{(\lambda)}_k(x) \\[0.3em]
(u_2)^{(\lambda)}_k(m,t) \:\phi^{(\lambda)}_k(x) \end{pmatrix} , \]
and similarly for~$\tilde{\psi}$. Choosing the spatial wave functions~$\phi^{(\lambda)}_k$ to
be orthonormal, we can apply Lemma~\ref{lemmamode} to each mode to obtain
\[ | \bra \p \psi | \p \tilde{\psi} \ket | \leq \pi 
\sum_{|\lambda| < |\Lambda|} \sum_{k=1}^{K(\lambda)} \sum_{s=\pm} \int_I
\big\| (f^s)^{(\lambda)}_k(m) \big\|\: \big\| (\tilde{f}^s)^{(\lambda)}_k(m) \big\|\: dm  \:. \]
Using current conservation~\eqref{currcons}, we can compute the norms of~$u^{(\lambda)}_k$
and~$\tilde{u}^{(\lambda)}_k$ asymptotically as~$t \rightarrow \pm \infty$
with the help of Lemma~\ref{lemma31}. This gives
\[ | \bra \p \psi | \p \tilde{\psi} \ket | \leq 2 \pi 
\sum_{|\lambda| < |\Lambda|} \sum_{k=1}^{K(\lambda)} \int_I
\big\| u^{(\lambda)}_k(m) \big\|\: \big\| \tilde{u}^{(\lambda)}_k(m) \big\|\: 
dm  \:. \]
Applying the Schwarz inequality gives the result.
\QED

We finally explain what the result of Lemma~\ref{lemmamode} means for the
decomposition of the solution space. Comparing~\eqref{modeform} with~\eqref{Smdef},
one sees that now the spectral subspaces of the fermionic signature operator~$\Sig_m$
no longer coincide with the solutions of positive and negative frequency.
This is also clear because in the time-dependent setting of the de Sitter space-time, the ``frequency''
of a solution is only defined asymptotically as~$t \rightarrow \pm \infty$, but not globally or at
intermediate times. Instead, the sum over~$s$ in~\eqref{psirep} corresponds to the fact
that we must take a suitable ``interpolation'' of the frequency splittings as experienced by
observers at asymptotic times~$t \rightarrow \pm \infty$.
Here the notion of ``interpolation'' can be understood similar as explained in~\cite[Section~5]{sea}
and~\cite[Section~6]{finite}.

\Thanks{{{\em{Acknowledgments:}}
We would like to thank Christian R\"oken for helpful discussions and comments. We are grateful
to Andreas Platzer for helpful comments on the manuscript.}

\providecommand{\bysame}{\leavevmode\hbox to3em{\hrulefill}\thinspace}
\providecommand{\MR}{\relax\ifhmode\unskip\space\fi MR }
\providecommand{\MRhref}[2]{%
  \href{http://www.ams.org/mathscinet-getitem?mr=#1}{#2}
}
\providecommand{\href}[2]{#2}

\end{document}